\documentclass[a4paper]{article}
\usepackage{victor}
\usepackage{booktabs}
\usepackage{pgfplotstable}
\pgfplotstableset{
every head row/.style={before row=\toprule,after row=\midrule},
every last row/.style={after row=\bottomrule},
col sep=semicolon}
\pgfplotstableread{figures/SI/country_code.csv}\datatable
\usepackage{longtable}
\usepackage[tableposition=below]{caption}
\title{Processes analogous to ecological interactions and dispersal shape the dynamics of economic activities}
\author{Victor Boussange$^{1, 2, \star}$,
Didier Sornette,
Heike Lischke$^{3}$,
and
Lo{\"i}c Pellissier$^{1,2}$\bigskip
\\
\begin{flushleft}
$^1$ \small{Swiss Federal Research Institute WSL, CH-8903 Birmensdorf, Switzerland}\smallskip\\
$^2$ \small{Ecosystems and Landscape Evolution, Institute of Terrestrial Ecosystems, Department of Environmental System Science, ETH Zürich, CH-8092 Zürich, Switzerland}\smallskip\\
$^3$ \small{Dynamic Macroecology, Land Change Science, Swiss Federal Research Institute WSL, CH-8903 Birmensdorf, Switzerland}\smallskip\\
$^\star$ Corresponding author, email: \href{mailto:bvictor@ethz.ch (VB)}{\texttt{bvictor@ethz.ch}}\smallskip
\end{flushleft}
}

\begin{document}
\maketitle

\graphicspath{{figures/}}





\begin{abstract} 
  The processes of ecological interactions, dispersal and mutations shape the dynamics of biological communities, and analogous eco-evolutionary processes acting upon economic entities have been proposed to explain economic change. This hypothesis is compelling because it explains economic change through endogenous mechanisms, but it has not been quantitatively tested at the global economy level.
  Here, we use an inverse modelling technique and 59 years of economic data covering 77 countries to test whether the collective dynamics of national economic activities can be characterised by eco-evolutionary processes.
  We estimate the statistical support of dynamic community models in which the dynamics of economic activities are coupled with positive and negative interactions between the activities, the spatial dispersal of the activities, and their transformations into other economic activities.
  We find strong support for the models capturing positive interactions between economic activities and spatial dispersal of the activities across countries. These results suggest that processes akin to those occuring in ecosystems play a significant role in the dynamics of economic systems.
  The strength-of-evidence obtained for each model varies across countries and may be caused by differences in the distance between countries, specific institutional contexts, and historical contingencies.
  Overall, our study provides a new quantitative, biologically inspired framework to study the forces shaping economic change.
\end{abstract}
\tableofcontents
\break

\section{Introduction}

  The fields of evolutionary biology and economics have exchanged ideas for the past two centuries \cite{Dopfer2007}, and analogies between processes in biology and economics have been increasingly used during the 20th century to explain economic change \cite{Ruth1996}.
  A pioneer was Joseph Schumpeter, who famously proposed that economic dynamics are driven by innovations that transform the economy -- so called periods of "creative destruction" \cite{schumpeter2017theory} -- similar to the punctuated equilibrium changes observed throughout the development of life on Earth \cite{gould1972}.
  The analogy with biology was further developed within the field of evolutionary economics \cite{Hodgson2019}, largely promoted by the seminal work of Richard Nelson \cite{nelson1985evolutionary}.
  The premise of evolutionary economics considers habits, customs and organisational routines as "replicators", i.e. atomic units playing the role of genes and defining, as a whole, the cohesive identity of an economic entity \cite{Hodgson2019}.
  Under this framework, the business strategies that firms apply determine how such firms transform commodities and knowledge into new knowledge, new technologies or other industrial products with added value.
  The replicators define the fitness of an economic entity within a given economic context, growing in terms of assets and human capital, surviving within an ecosystem of entities \cite{Hodgson2002}, and continuously adapting and experiencing evolutionary processes \cite{Veblen1898}.
  While biological analogies have resulted in useful insights into the plausible drivers of economic growth \cite{Dopfer2007}, most investigations of the proposed pathways have been qualitative.
  However, computational tools have recently made it possible to test biological hypotheses against data to gain a quantitative understanding of the ecological and evolutionary processes shaping the dynamics of ecological systems \cite{Pontarp2019,Boussange2022a,Skeels2022}. These tools could leverage the qualitative insights gained from biological analogies and provide a quantitative framework to investigate the forces shaping economic change.
  
  Interactions between biological organisms, dispersal (movement of individuals across space), and mutations are fundamental processes that drive the dynamics of ecosystems \cite{Vellend2010}, and similar eco-evolutionary processes may shape the dynamics of economic systems.
  Analogous to biological organisms, economic entities interact in a mutualistic or competitive fashion \cite{Pistorius1997}. While biological organisms engage in negative interactions, e.g. when competing for similar resources \cite{GRIME1973}, economic entities have negative effects on each other when they rely on workforces with similar knowledge or when they attract similar potential investors \cite{Wernerfelt1989}.
  Positive interactions are observed between biological organisms through e.g. accumulation of nutrients, provision of shade, or protection from herbivores \cite{Callaway2002}. In parallel, positive effects between economic entities appear when they are connected through supply chains \cite{Ozman2009,Saavedra2009a}, or when they benefit from positive agglomeration externalities \cite{VanDerPanne2004}, such as knowledge spillovers \cite{Caragliu2016} or the attraction of elite workers \cite{Cohendet2018}.
  Dispersal processes play a major role in the development of ecosystems, e.g. when colonisation initiates ecological succession through the settlement and growth of new species \cite{Leibold2004}. Similarly, socio-economic processes contribute to the diffusion of knowledge and organisational routines across space. These processes involve e.g. international business expansions \cite{Andersen1993,Zahra2000,Luo2007}, labour mobility \cite{Boschma2008}, and the diffusion of innovations through social networks \cite{RogersEverettM2003DoI,Keller2004,Bahar2014a}.
  %
  Finally, evolutionary processes allow organisms to adapt to changing environmental conditions \cite{Bell2017}, while in economic systems, variations in organisational routines allow economic entities to adapt to economic contexts \cite{Cordes2006} and transform into new economic entities \cite{Freeman2002,Hodgson2004,Aldrich2008}. 
  While processes acting upon economic activities, including interactions between the activities, their spatial dispersal, and their transformations into other economic activities, are regularly documented in evolutionary economics, the significance of these processes regarding long-term economic change has seldom been quantified.

  Focusing on endogenous forces and borrowing concepts and methods from biology, a number of modelling approaches have broken with the traditions of standard economic modelling \cite{10.1093/cje/bet027} to study the processes driving economic patterns \cite{Tacchella2018}.
  For instance, \cite{Saavedra2009a} investigated the effect of cooperation within firms with a model of consumer--resource interactions. Moreover, \cite{Scholl2020} developed a theory of market ecology to interpret market phenomena and predict market behaviour, where financial trading strategies are considered analogous to biological species.
  Lotka--Volterra models have been used to predict technology evolution \cite{Zhang2018}, to evaluate competition between products and firms \cite{Modis1997,Saavedra2014}, to understand the drivers of market share dynamics \cite{Farmer1999,Michalakelis2011,Marasco2016,Gatabazi2019}, and to estimate the value of firms \cite{Cauwels56}.
  Further, \cite{Applegate2021} investigated an ecological model of competition--colonisation dynamics to understand the distribution of firm sizes, and \cite{Suweis2015} used a dynamic community model to evaluate the effect of international trade on global food security.
  %
  %
  Fine-grained datasets on economic activities, together with dimensionality reduction techniques, have provided insights into the endogenous processes shaping economic development \cite{Mealy2019,Hidalgo2021}.
  In particular, the detailed global trade data compiled by the United Nations Statistical Division and cleaned by \cite{Hidalgo2021} consists of a time series of 59 years of economic activity that can be combined with inverse modelling techniques to learn about the processes influencing the long-term dynamics of economic systems.
  
  Here, we investigate whether eco-evolutionary processes can quantitatively explain economic growth, using an inverse modelling technique together with 59 years of data on economic activity. By relating the temporal evolution of the capital of economic activities to the temporal evolution of the biomass of functional groups within an ecosystem, we quantitatively estimate the effects of eco-evolutionary processes on the long-term development of national economies.
  Functional groups, as the aggregation of species sharing similar characteristics and having similar functions within an ecosystem, are the elemental units required to model ecosystem dynamics. Analogously, as the aggregation of firms with similar outputs \cite{Applegate2021}, we consider nine economic activities to form the elemental units of economic systems, and we conduct our investigation by modelling their temporal dynamics.
  We consider a null model implementing the fundamental processes of self-replication and self-limitation ($\modnull$), where no couplings between economic activities are captured. We contrast $\modnull$ with alternative dynamic community models coupling the dynamics between economic activities, including both negative and positive ecological interactions between the activities ($\modalphan,\modalphap$), the spatial dispersal of the activities ($\moddelta$), and their transformations into other economic activities ($\modmu$).
  We use the machine learning framework of \cite{Boussange2022a} to estimate the maximum likelihood of each model for 96 countries. We use a model selection technique to evaluate the statistical support for each alternative model, based on a dataset of national exports from 1962 to 2020, taken as a proxy for the temporal development of the capital of economic activities. 
  We first show that the machine learning method, together with the model selection procedure, can provide support for the generating processes in a controlled experiment.
  By applying the model selection procedure to the empirical data, we then find evidence for eco-evolutionary processes and observe differences in model support across countries, which we relate to socio-economic drivers. 
  Our study provides a perspective on the drivers of economic dynamics that complements mainstream economic theory, and it serves as the basis for a biological understanding of the endogenous forces determining economic growth.

\section{Methods}\label{sec:methods}

\subsection{Eco-evolutionary model to characterise the dynamics of economic activities}

We derive a general dynamic community model, where the dynamics of national economic activities are driven by self-replication and self-limitation of the activities, by interactions between the activities, by the spatial dispersal of the activities across countries, and by transformations of the activities into other economic activities within the countries. The general model, presented below, is subsequently split into alternative sub-models to test the support for each process.

\begin{figure}[ht]
  \includegraphics[width=0.8\textwidth]{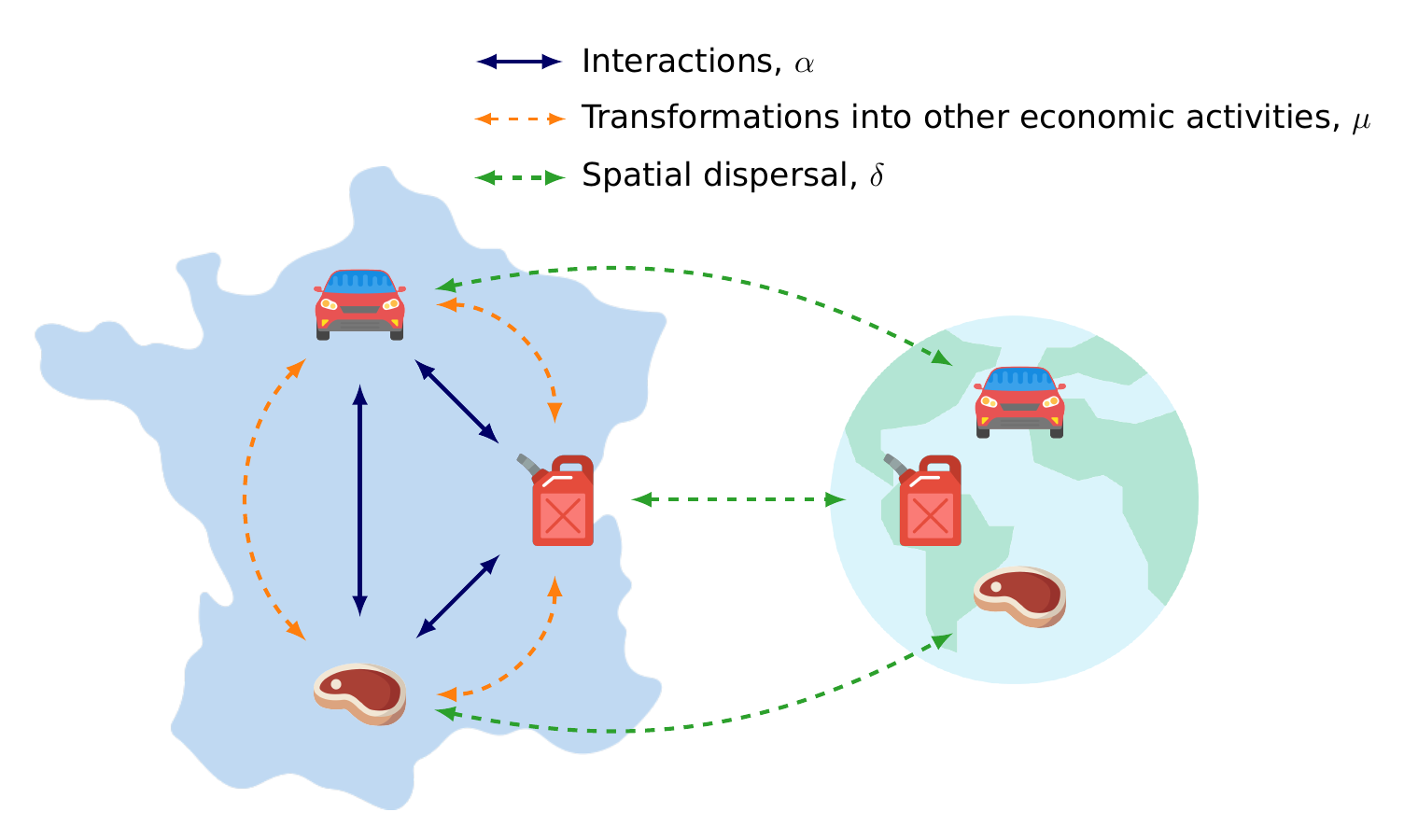}
  \centering
  \caption{\small
      \textbf{General dynamic community model used to characterise the dynamics of economic activities, capturing interactions between the activities, the spatial dispersal of the activities, and their transformations into other economic activities.}
      For each country $c$, each one of the nine economic activities considered is assumed to be driven by replication and self-limitation processes, characterised by a growth rate $r_i\hc$ and a self-limitation coefficient $b_i\hc$, respectively, with the values assumed to be country-dependent and determined by the institutional context and the available resources (in terms of human capital, knowledge capital and natural resources). 
      In addition, interactions between activities within each country are determined by $\alpha\hc$ and can be mutualistic ($\alpha\hc > 0$) or competitive ($\alpha\hc < 0$).
      Due to spatial dispersal, the capital of an economic activity can experience positive (resp. negative) fluxes of capital when it is lower (resp. higher) than the average global capital for the economic activity at time $t$, arising at rate $\delta\hc$.
      Due to transformations into other economic activities, the capital of an economic activity can experience positive (resp. negative) fluxes of capital when it is lower (resp. higher) than the capital of other economic sectors within the country at time $t$, arising at rate $\mu\hc$.
      The parameters $r_i\hc$, $b_i\hc$, $\alpha\hc$, $\delta\hc$ and $\mu\hc$ that best fit the empirical data are estimated for each country for each sub-model in \cref{eq:submodels1,eq:submodels2,eq:submodels3,eq:submodels4,eq:submodels5}, to investigate whether the proposed processes have a significant effect on the temporal dynamics of economic activities. This figure was designed using resources from Flaticon.com.
      }\label{fig:model}
\end{figure}

In the general dynamic community model, which includes all the investigated processes, the rate of change of the size of an economic activity $i$ in country $c$, denoted as $n_i\hc $, is determined as follows:

\begin{equation}\label{eq:model_general}
  \begin{split}
    \tfrac{d}{dt} n_i\hc(t)  &= r_i\hc  n_i\hc(t) \left(1 -   b_i\hc  n_i\hc(t)  + \sum_{j \neq i}^{N\hc} \alpha\hc _{i,j} n_j\hc(t) \right) + \sum_l^M \delta^{(l,c)}_{i} \left( n_i^{(l)}(t) - n_i\hc(t)  \right) \\
    &\qquad + \sum_j^{N\hc} \mu\hc _{j,i}\left( n_j\hc(t)  - n_i\hc(t)  \right),
  \end{split}
\end{equation}
where $N\hc$ is the number of economic activities considered in country $c$ and $M$ is the number of countries considered.
In \cref{eq:model_general}, the first summand corresponds to a Lotka--Volterra model for biological communities (e.g. \cite{Bunin2017,Scheffer2006a,Case1990}), where  $r_i\hc $ is the growth rate, involved in self-replication, and $b_i\hc $ is the self-interaction coefficient, involved in self-limitation. $b_i\hc $ can also be interpreted as the inverse of the country's carrying capacity for the activity. 
The individual dynamics of economic activities are altered by interactions, where $\alpha_{i,j}\hc $ captures the interaction between activity $ i $ and activity $ j $ at location $c$. The interaction between $i$ and $j$ is mutualistic if $\alpha_{i,j}\hc > 0$, and competitive if $\alpha_{i,j}\hc < 0$, altering the rate of change of the activity $i$ positively or negatively, respectively, by the presence of other activities.
The second summand accounts for spatial dispersal across locations (e.g. \cite{Tilman1994b}), where $\delta^{(l,c)}_{i}$ is the rate of spatial dispersal for activity $i$ between locations $l$ and $c$.
The last summand corresponds to economic activity transformations within country $c$, where $\mu\hc _{j,i}$ is the rate of transfer dictating how fast activity $j$ transforms into activity $i$. In the sense of a quasi-species model \cite{eigen1988molecular}, this term can be interpreted as accounting for evolutionary processes, and can contribute to the development of activity $i$ in the presence of activity $j$.
A graphical representation of the model is given in \cref{fig:model}.

\Cref{eq:model_general} involves a high number of independent parameters to fit (i.e. $N\hc(1 + M + 2N\hc)$ parameters for each country), which can be reduced for the sake of parsimony under mean field assumptions. Specifically, we assume that all economic activities interact similarly, so that $\alpha_{i,j}\hc  = \alpha\hc $. We further assume that activity transformations are symmetric and occur at similar rates for all activities, so that $\mu_{i,j}\hc  = \mu\hc $, and we assume that spatial dispersal occurs at a similar rate for all activities and all countries, so that $\delta^{(l,c)} _{i} = \delta^{(c)}$. These mean field assumptions reduce the number of parameters to $2N+3$. 
In summary, we assume that the parameters $r_i\hc$ and $b_i\hc$ are activity- and country-dependent, and that $\alpha\hc$, $\mu\hc$ and $\delta\hc$ are country-dependent. We further assume that the values of these parameters are determined by the activity's characteristics and the country's institutional system (e.g.~taxation regime, system of innovation, legal system, intellectual property rights, and socio-cultural background) and resources (e.g.~labour force, knowledge capital, agricultural resources, mineral resources and energy resources), but that they do not change through time.

To investigate how the data supports each of the processes embedded in \cref{eq:model_general}, we further decompose the model into five sub-models, as follows:

  \begin{align}
  \label{eq:submodels1}
    \M_{null}&:  &&\tfrac{d}{dt} n_i\hc(t)  = r_i\hc  n_i\hc(t) (1 -  b_i\hc n_i\hc (t) ) \\
  \label{eq:submodels2}
    \modalphap&: && \tfrac{d}{dt} n_i\hc(t)  = r_i\hc  n_i\hc(t) \left(1 -  b_i\hc n_i\hc(t) + \alpha\hc  \sum_{j \neq i} n_j\hc(t) \right), \quad \alpha > 0\\
      \label{eq:submodels3}
    \modalphan&: && \tfrac{d}{dt} n_i\hc(t)  = r_i\hc  n_i\hc(t) \left(1 -  b_i\hc  n_i\hc(t) + \alpha\hc  \sum_{j \neq i} n_j\hc(t) \right), \quad \alpha < 0\\
      \label{eq:submodels4}
    \moddelta&: && \tfrac{d}{dt} n_i\hc(t)  = r_i\hc  n_i\hc(t) (1 - b_i\hc  n_i\hc(t) ) + \delta\hc \left(\overline{n_i\hc}(t) - n_i\hc(t) \right) \\
      \label{eq:submodels5}
    \M_{\mu}&: && \tfrac{d}{dt} n_i\hc(t)  = r_i\hc  n_i\hc(t) (1 -  b_i\hc n_i\hc(t) ) + \mu\hc \sum_j^{N\hc} \left( n_j\hc  (t) - n_i\hc(t)  \right)
  \end{align}

where 
\begin{equation}\label{eq:n_overline}
      \overline{n_i\hc } = \frac{1}{M-1} \sum_{l \neq c}^M n_{i}^{(l)}
\end{equation}
accounts for the capital of activity $i$ at the global level, excluding country $c$. The term $\overline{n_i\hc }$ appearing in \cref{eq:submodels5,eq:n_overline} arises from the mean field assumption used for spatial dispersal, where 

\begin{equation}
  \begin{split}
    \sum_l^M \delta^{(l,c)}_{i} \left( n_i^{(l)}(t) - n_i\hc(t)  \right) &= \delta\hc  \sum_l^M  \left(n_i^{(l)}(t) - n_i\hc(t) \right) \\
    &= \delta\hc \sum_{l \neq c}^M \left(n_i^{(l)}(t) - n_i\hc(t) \right) \\
    &= \delta\hc \left( \sum_{l \neq c}^M n_i^{(l)}(t) - (M - 1) n_i\hc \right) \\
    &\propto \delta\hc \left( \overline{n_i\hc } - n_i\hc \right)
  \end{split}
\end{equation}

The simplest model $\modnull$ embeds self-replication and self-limitation without any further coupling forces acting upon economic activities, and is considered the null model.
The alternative models $\modalphap$, $\modalphan$, $\moddelta$, and $\modmu$ capture self-replication and self-limitation together with one of the additional eco-evolutionary processes investigated.
Because we find inconsistent maximum likelihood estimations in the controlled experiment detailed in the \nameref{sec:results} for models incorporating combinations of eco-evolutionary processes, we do not test more complex models.
In the following, we use $\M(t,\theta\hc )$ to designate the vector of economic activity capitals predicted by model $\M$ at time $t$ for country $c$ with the parameter vector $\theta\hc$, which includes the growth rates $r_i\hc $, the self-limitation rates $b_i\hc$, the specific model parameters $\alpha\hc$, $\mu\hc$ and $\delta\hc$, and the initial condition vector $n\hc (t_0)$.


\FloatBarrier

\subsection{Empirical data and model likelihood}

\begin{figure}
  \centering
  \includegraphics{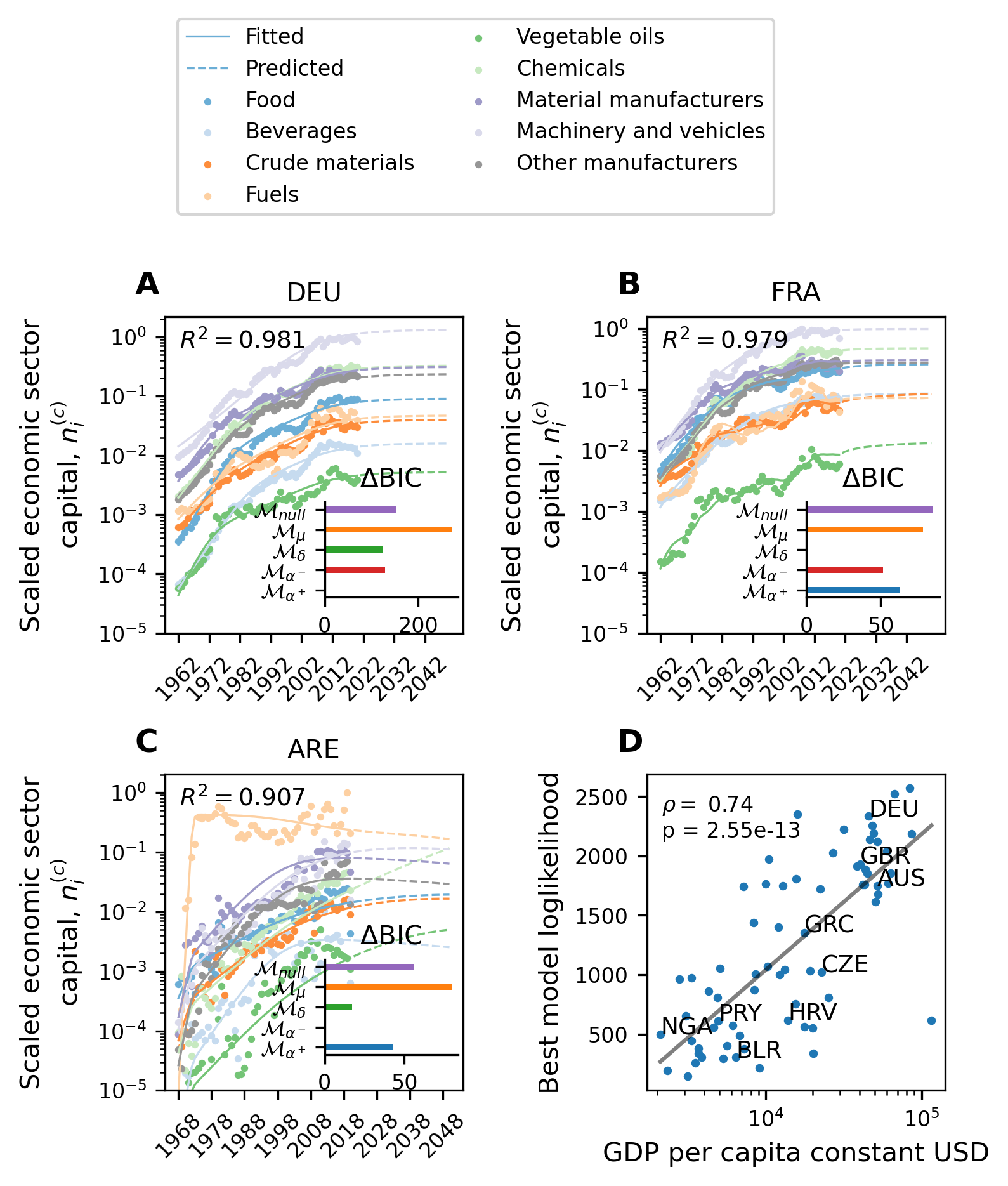}
  \caption{\small \textbf{Best model fits and associated predictions for Germany (DEU), France (FRA) and United Arab Emirates (ARE), and effect of gross domestic product (GDP) on the best model log-likelihoods.}
    \textbf{A}--\textbf{C} Best model fits and associated predictions for DEU, FRA and ARE. Dots represent the empirical data, solid lines correspond to the best model fits, and dashed lines correspond to the best model predictions. The inset plots display the relative Bayesian information criterion score ($\Delta \BIC$) for each country.
    \textbf{D} Effect of GDP on model log-likelihood. The solid line corresponds to a linear fit.
  }\label{fig:fits}
\end{figure}

We consider time series of global trade data as a proxy for the evolution of the capital of economic activities through time. Export data is synthetic of the capital of an economic activity within a country; the more a country exports the output of a given economic activity, the more competitive the economic activity is in the global market, and therefore the larger the economic activity is in terms of capital \cite{Tacchella2018}. Specifically, we use the database compiled by the United Nations Statistical Division COMTRADE, categorised in the Standard International Trade Classification (SITC, revision 2) at the 1 digit level, covering 11 categories of economic activities in 249 countries from 1962 to 2020. Because of their low quality \cite{Hidalgo2021}, we discard the "unspecified" and "services" categories, resulting in nine economic categories considered in the models (see \cref{fig:fits} for details). As shown in the \nameref{sec:results}, this large number of data points contains the information necessary to recover the plausible generating processes.
Export values for activity $i$ in country $c$, denoted by $X_{i}\hc (t)$, are discounted by the national population at time $t$, denoted by $P\hc (t)$, in order to compare the economic variables across time and across countries. Consequently, the observation data for the capital of activity $i$ in country $c$ is calculated as
$ y_i\hc (t) = X_{i}\hc (t) / P\hc (t) $.
%
In each country, we only consider economic activities that have sustained a significant size relative to the world trends for at least 4 years (revealed comparative advantage $>1$; see \cite{Hidalgo2021}).
We assume that the observation data is contaminated by white noise with log-normal distribution $\epsilon$ with zero mean and a variance--covariance matrix $\Sigma = \sigma^2 I$, where $\sigma$ is the noise level, which is a reasonable error model for population dynamics \cite{Schartau2017}, so that
$y\hc _i(t)  = n\hc _i(t) \exp(\epsilon_i\hc (t))$.
As a result, we express the likelihood of model $\M$ in country $c$, denoted by $\LL(\theta\hc_\M | {\by\hc }, \M)$, as:

\begin{equation}\label{eq:likelihood_lognormal}
  \LL(\theta\hc_\M | {\by\hc }, \M) = \prod_{j=1}^{T\hc} \frac{1}{\sqrt{(2\pi)^{N\hc}|\Sigma|} y\hc(t_j)} \exp \left(-\frac{1}{2} \Big[ \tilde{d}\hc(t_j)\Big]^{T\hc} \Sigma^{-1} \Big[ \tilde{d}\hc(t_j)\Big] \right)
\end{equation}
where $\theta\hc_\M$ refers to the parameters and the initial conditions for model $\M$; $\by\hc = (y\hc(t_1), \allowbreak \dots, y\hc(t_T))$  designate the $T\hc$ time points of economic data available for country $c$; $ y\hc(t) = (y_1(t), \dots, y\hc_N(t_T))$ is the vector of economic activity capital at time $t$; and $\tilde{d}\hc(t_j) = \ln(y\hc(t_j)) - \ln(\M(t_j,\theta\hc_\M))$. 
In the next section, we describe how we obtain the maximum likelihood estimate $\hat\theta\hc_\M$ that maximises \cref{eq:likelihood_lognormal}, given $\by\hc$, for each model $\M$ and country $c$. $\hat\theta\hc_\M$ is used to select the most probable model, given the data \cite{Burnham2002}. In the following, we drop the indices $\hc$ for clarification.

\FloatBarrier

\subsection{Maximum likelihood estimation}\label{parameter-inference}

To obtain the maximum likelihood of each model we employ the machine learning framework detailed in \cite{Boussange2022a}, which is based on a segmentation method combined with automatic differentiation and optimisers commonly used in deep learning. The likelihood of model $\M_i$ is maximised by training $\M_i$ against segments of data comprising only $K < T\hc$ data points of the full time series, where the parameters and the initial conditions for each segment are estimated.
The segmentation method ensures convergence towards the maximum likelihood estimate, provided that the choice of $K$ is appropriate, given the data and the model investigated \cite{Boussange2022a}. 
A large $K$ might induce convergence towards a local minimum, while a low $K$ might flatten the likelihood landscape, where all models would be assigned equal support. 
In the controlled experiments presented in the \nameref{sec:results}, we find that $K=20$ (models trained against data segments of 20 years) ensures good convergence, while maintaining reasonable discrimination between the models (\cref{fig:synthetic_test_all_AIC}). We therefore use $K=20$ throughout all the experiments detailed in the manuscript, and discard countries where data is available for less than $K = 20$ years.
We employ the Julia package \textbf{PiecewiseInference.jl} \cite{Boussange2022a} for the numerical implementation. We use the gradient descent optimiser ADAM \cite{Kingma2014} during the first 800 epochs to converge in the basin of attraction of the maximum likelihood estimate. We substitute the Broyden-Fletcher-Goldfarb-Shanno optimiser (BFGS) \cite{fletcher2013practical} for ADAM for the final 800 training epochs to ensure faster and more accurate convergence.
As a cross-checking procedure, for each country and each model we perform five optimisation runs with different initial parameter values, where $r_i$, $b_i$, $\alpha$, $\mu$ and $\delta$ are drawn from the uniform random distributions $\mathcal{U}_{[0.05, 0.15]}$, $\mathcal{U}_{[0.5,1.5]}$, $\mathcal{U}_{[0.5,1.5]}$, $\mathcal{U}_{[0.0005, 0.0015]}$ and $\mathcal{U}_{[0.0005, 0.0015]}$, respectively.
We then take the best run among the five optimisation runs, making sure that the likelihood estimates from each run are similar (see \cref{tab:results} for full details of the results).
%


\section{Results}
\label{sec:results}
\subsection{Validation with synthetic data}\label{sec:synthetic}
\begin{figure}
  \centering
  \includegraphics{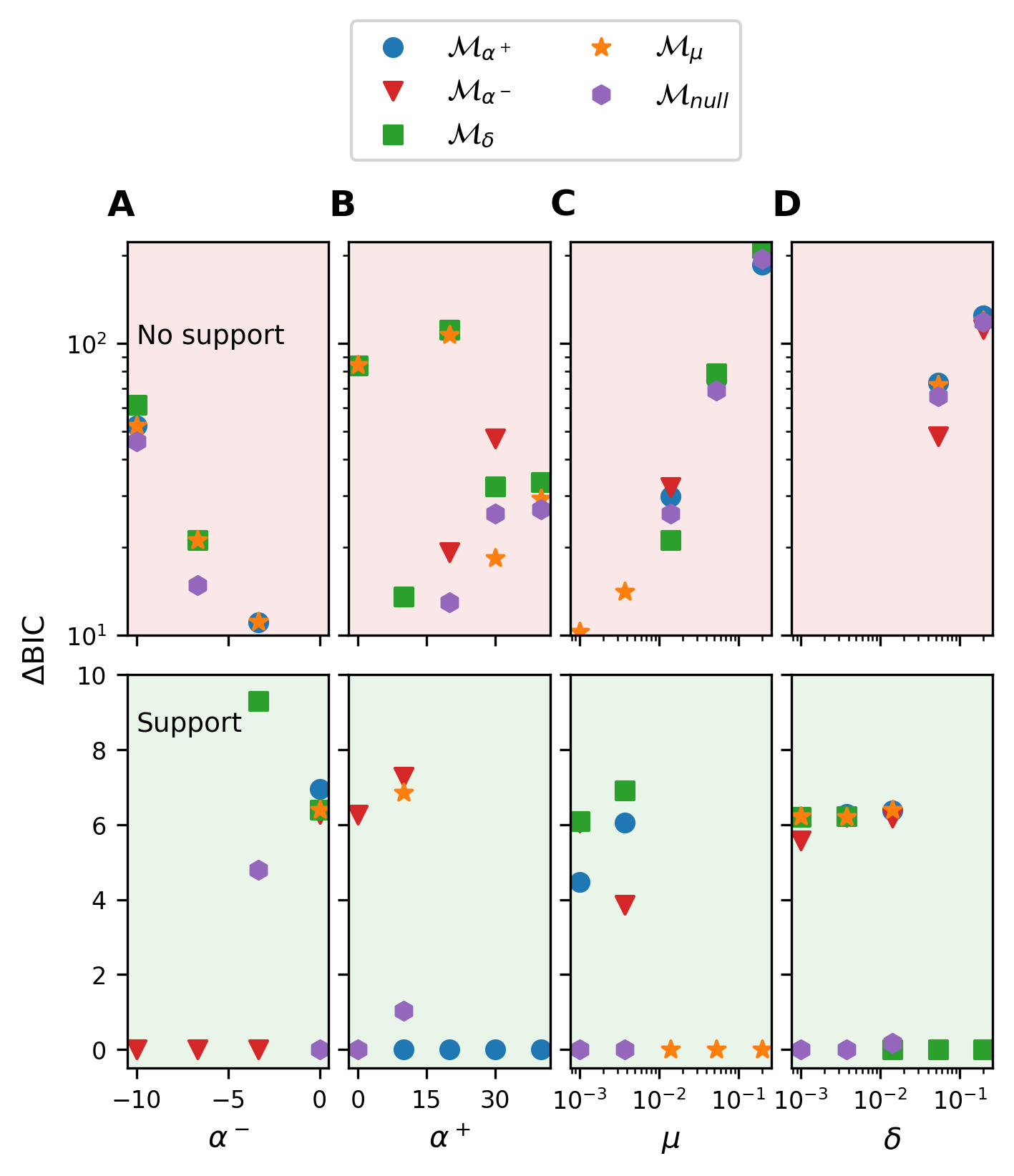}
  \caption{\small
  \textbf{Validation of the model selection procedure.} In \textbf{A} and $\textbf{B}$ we generate data with models $\modalphan$ and $\modalphap$ by varying the value of $\alpha$. In \textbf{C} we generate data with model $\moddelta$ by varying the value of $\delta$, and in \textbf{D} we generate data with model $\modmu$ by varying the value of $\mu$. We then contaminate the generated data with realistic noise ($\sigma = 0.2$), and determine the maximum likelihood of all models $\modnull$, $\modalphan$, $\modalphap$, $\moddelta$ and $\modmu$. The maximum likelihood of each model is then used to compute the relative Bayesian information criterion score (BIC) for each model, $\Delta \BIC$. Comparing $\Delta \BIC$ provides a strength-of-evidence for each model. 
  \textbf{A}--\textbf{D} show that the model selection procedure is valid, as only the true models are placed in the "support" category when no other model is given support (equivalently, when all other models are placed in the "no support" category).}
  \label{fig:synthetic_test_all_AIC}
\end{figure}

We first investigate whether the inverse modelling technique of \cite{Boussange2022a}, together with the proposed sub-models in \cref{eq:submodels1,eq:submodels2,eq:submodels3,eq:submodels4,eq:submodels5}, can detect signatures of eco-evolutionary processes in a controlled experiment.
We proceed by generating multiple synthetic datasets from the models $\modalphan$, $\modalphap$, $\M_\delta$ and $\modmu$ with realistic $r_i$ and $b_i$ parameters, with different values for the parameters $\alpha$, $\mu$ and $\delta$, and with different values for the noise level $\sigma$. We consider $\modnull$, $\modalphan$, $\modalphap$, $\M_\delta$ and $\modmu$ as equally plausible candidate models for each generated dataset, and we apply the maximum likelihood estimation method detailed in the \nameref{sec:methods} to obtain the maximum likelihood $\LL(\hat \theta, \M | \by)$ of each model for each generated dataset. We then use the Bayesian information criterion (BIC) to select the model with the strongest strength-of-evidence in relation to the data \cite{Mangan2017}, calculating the BIC for model $\M_i$ as $\BIC_{\M_i} = -2 \ln(\LL(\hat \theta, \M_i | \by)) + k_{\M} \ln(N T ) $,
where $N$ is the number of activities in country $c$, $T$ is the number of time points in the time series considered ($N T$ therefore being the number of data points for country $c$), and $k_{\M}$ is the number of parameters in the model $\M$ for country $c$.
The BIC ranks the most probable models by penalising complexity to balance information loss and parsimony, where candidate models with the lowest scores are ranked as the most likely \cite{Mangan2017}.
We consider the relative BIC score $\Delta \BIC_{\M_i}$, which allows a strength-of-evidence comparison across models and is calculated as 
$\Delta \BIC_{\M_i} = \BIC_{\M_i} - \min_{j} \BIC_{\M_j}$. 
We expect the $\Delta \BIC$ scores to only provide support for the true generating model when the process considered has a significant effect on the observed dynamics.
Under realistic observational noise ($\sigma = 0.2$), we find overall strong support for the true models (i.e. when $\M_i$ is the true generating model, $\Delta\BIC_{\M_{i}} = 0$ and $\Delta\BIC_{\M_{j,j\neq i}} > 10$; \cref{fig:synthetic_test_all_AIC}\textbf{A--C}) when the values of the parameters $\alpha$, $\mu$ and $\delta$ exceed a certain threshold. 
%
%
%
Moreover, as the values of $\alpha$, $\mu$, and $\delta$ become more extreme, more support is given to the true generating model (\cref{fig:synthetic_test_all_AIC}\textbf{A--C}).
In contrast, when the signature of the process underlying the dynamics is not sufficiently strong, the null model $\modnull$ is given the most strength-of-evidence ($\Delta\BIC_{\modnull} = 0)$.
%
%
These results hold for varying noise levels (see \cref{figSI:synthetic_test_all_AIC} for $\sigma = 0.3$), indicating that the models, together with the maximum likelihood estimation method, are well adapted to investigate the influence of eco-evolutionary processes with the dataset considered.
Based on this experiment and classical model selection criteria \cite{Burnham2002}, we accept the hypothesis that the null model is the best model when $\Delta \BIC_{\modnull} \leq 10$. In addition, we conclude that $\M_i$ is supported against $\M_j$ when $\BIC_{\M_i} - \BIC_{\M_j} < -10$. Finally, we conclude that $\M_i$ is the best model if $\Delta \BIC_{\M_i} = 0 $ and if for all other models $\M_{j}$, $j \neq i$ we have that $\Delta \BIC_{\M_{j}} > 10$.
Overall, the proposed eco-evolutionary model, together with maximum likelihood estimation, leads to a good discrimination ability when combined with the BIC-based model selection procedure, and can provide strength-of-evidence for eco-evolutionary processes that may shape the dynamics of economic systems.

\FloatBarrier

\subsection{Discrepancies in dynamical regime across countries}

We apply the maximum likelihood estimation method detailed in the \nameref{sec:methods} to obtain the maximum likelihood of each model for the world's 96 richest countries (highest gross domestic product [GDP] per capita as of 2020, see list in \cref{tab:country_codes}) for which population and export data is available . 
We first investigate the quality of best fit, and then detail the results for each model in the following section. 
In the following, we refer to countries by their ISO 3166-1 reference code. Full country names are given in \cref{tab:country_codes}.
Details of the numerical simulations are provided in \cref{tab:results}.
Of the 96 countries investigated, only 78 have sufficiently long time series to be included in the analyses (countries with at least 20 years of data, see \nameref{sec:methods}), while 77 show consistent maximum likelihood estimates (best model coefficient of determination $R^2$ > 0).
Among these 77 countries, we find that the best models have a good fit to the data and capture the long-term growth of economic activities (median explained variance $R^2 = 0.940$; see \cref{fig:fits} and \cref{figSI:fit_alphan,figSI:fit_alphap,figSI:fit_delta,figSI:fit_mu} for graphical illustrations of the fit).
While this demonstrates the relevance of the proposed models, we observe discrepancies across the countries in how well the best models perform (standard deviation of explained variance std$(R^2)= 0.074$; see \cref{fig:fits}\textbf{D} for a graphical illustration of the variance in terms of the model log-likelihoods).
%
%
This discrepancy is explained by the fact that the number of data points available differs among the countries, and that more data points yield a higher model log-likelihood ($\beta = 0.928 \pm 0.046$, $p<0.001$; \cref{table:gdp_loglikelihood}).
We additionally observe a positive effect of the GDP on the residuals ($\beta = 0.693 \pm 0.187$, $p<0.001$; \cref{table:gdp_loglikelihood} and \cref{fig:fits}\textbf{D}), indicating that the proposed models are more suited to characterise rich economies. The residuals are further negatively associated with the number of economic activities ($\beta = -0.230 \pm 0.071$, $p<0.01$; \cref{table:gdp_loglikelihood}), indicating that the proposed models might be less appropriate for quantifying the dynamics of diversified economies.
%
%

\subsection{Evidence for eco-evolutionary processes}

\begin{figure}
  \center
  \includegraphics{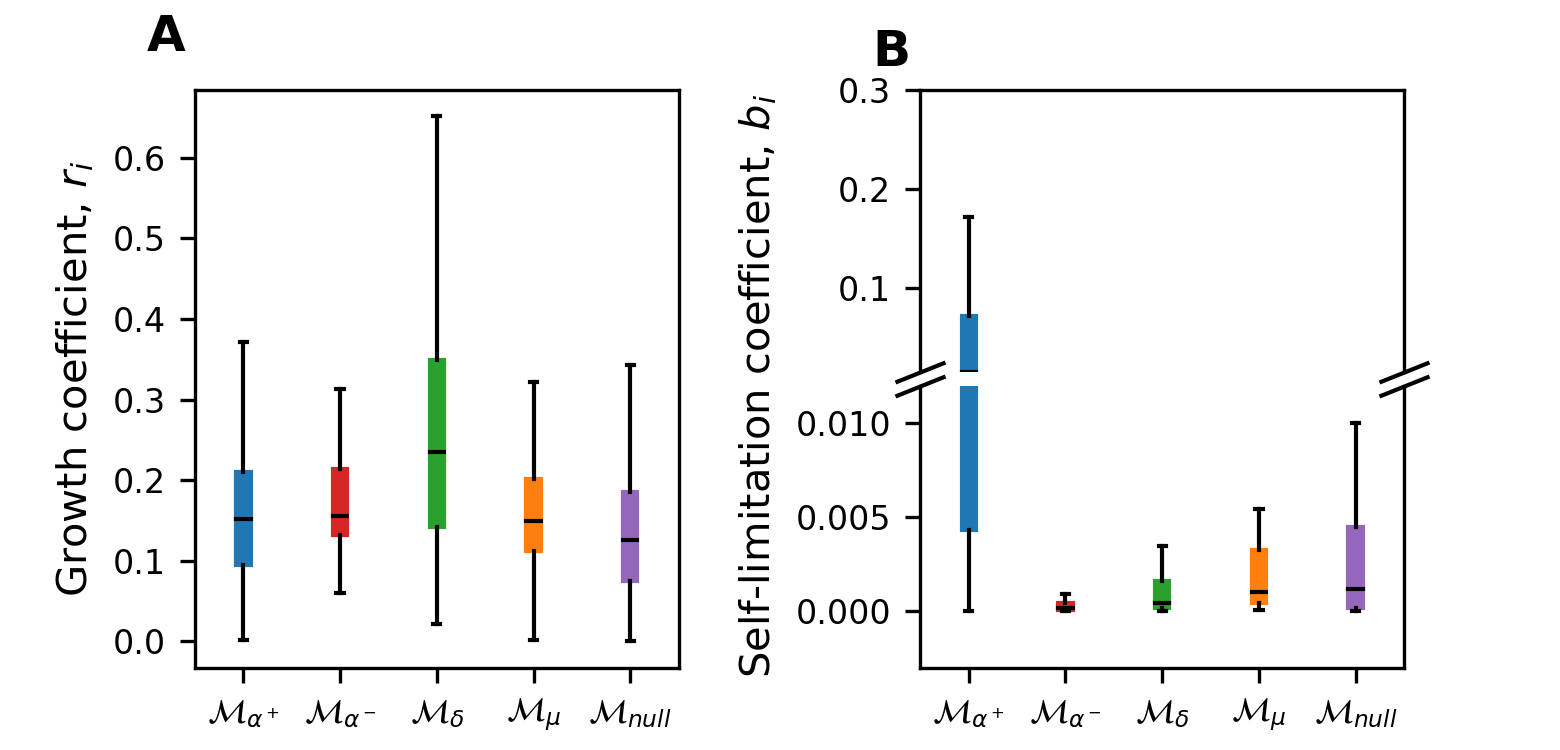}
  \caption{\small \textbf{Differences in growth rate $r_i$ and coefficient of self-limitation $b_i$ between $\modalphan$, $\modalphap$, $\moddelta$, $\modmu$, and $\modnull$}. In \textbf{A} and \textbf{B}, we aggregate the parameters $r_i$ and $b_i$ across all economic activities and all countries where models are supported. \textbf{A} shows that $\moddelta$ overestimates $r_i$ relative to the other models. This is because most economic activities experience a negative capital flux for a dispersal rate $\delta$ > 0, and consequently need to grow more to compensate for the losses. \textbf{B} shows that $\modalphap$ is associated with considerably larger $b_i$ values  relative to the other models. Such large values offset the beneficial effect of positive interactions on growth dynamics.
   }\label{fig:r_b}
\end{figure}

Applying the model selection procedure detailed in \cref{sec:synthetic}, we investigate in detail the effect of eco-evolutionary processes on the dynamics of the considered countries. We find that the null model is rejected in 45 of the 77 countries, in favour of alternative models (\cref{fig:aggregate_countries}\textbf{B}).
Among the alternative models, $\modalphap$, capturing positive interactions between economic activities, is the most frequently supported (\cref{fig:aggregate_countries}\textbf{B},\textbf{C}), and is given considerably more support against $\modnull$ than $\modalphan$ and $\modmu$ (\cref{fig:aggregate_countries}\textbf{A},\textbf{D}).
By comparing the parameter values across the models (\cref{fig:r_b}), we further find that $\modalphap$ is associated with an increase in the self-limitation coefficient $b_i$, which offsets the beneficial effect of positive interactions on growth dynamics (\cref{fig:r_b}\textbf{B}).
The model with spatial dispersal, $\moddelta$, is the second most supported model (\cref{fig:aggregate_countries}\textbf{B},\textbf{C}) and is given the most support against $\modnull$ (\cref{fig:aggregate_countries}\textbf{A},\textbf{D}).
$\moddelta$ is the only model that reproduces the oscillations observed in the empirical data (e.g. FRA in \cref{fig:fits}; BEL and GBR in \cref{figSI:fit_delta}). Oscillations arise from the term $\overline{n\hc}$ through mismatches between the local capital size and the global capital size of a given economic activity.
The models capturing negative interactions, $\modalphan$, and capturing economic activity transformations, $\modmu$, are the least supported and given the least strength-of-evidence (\cref{fig:aggregate_countries}\textbf{B},\textbf{D}), but they are still ranked as most supported models in some countries (\cref{fig:aggregate_countries}\textbf{A}). 
In contrast to $\modalphan$, $\moddelta$, $\modmu$ and $\modalphap$, $\moddelta$, $\modmu$, which implement complementary processes, $\modalphan$ and $\modalphap$ are structurally antagonistic, but are nonetheless equally supported in some countries (e.g. FIN and FRA in \cref{fig:aggregate_countries}\textbf{A}). This ambiguity highlights that the success of each model arises from its ability to capture different features of the empirical data.
%
Overall, positive interactions and spatial dispersal stand out as the most supported processes, in terms of both the number of countries where they are given support and the strength-of-evidence against $\modnull$.

\begin{figure}
  \center
  \includegraphics[height=0.85\textheight]{figure_aggregate_countriesv2_evidence_for_eco_evo.png}
  \caption{\small \textbf{Statistical support for the dynamic community models $\modalphan$, $\modalphap$, $\M_\delta$ and $\modmu$}. 
  \textbf{A} Strength-of-evidence comparison for $\modalphan$, $\modalphap$, $\moddelta$ and $\modmu$ against $\modnull$, for countries where $\modnull$ is rejected, presented in descending order. Full country names are given in \cref{tab:country_codes}. 
  \textbf{B} Number of countries where models $\modnull$, $\modalphap$, $\modalphan$, $\moddelta$, and $\modmu$ are ranked as the best models. 
  \textbf{C} Number of countries where models $\modalphan$, $\modalphap$, $\moddelta$ and $\modmu$ are supported against $\M_{null}$.
  \textbf{D} Global strength-of-evidence comparison for $\modalphan$, $\modalphap$, $\moddelta$ and $\modmu$ against $\modnull$.
  \textbf{A}--\textbf{D} show that $\modalphap$ and $\moddelta$ stand out as the most supported models, in terms of both the number of countries where they are given support and the strength-of-evidence against $\modnull$.
  }\label{fig:aggregate_countries}
\end{figure}
\FloatBarrier


\FloatBarrier

\section{Discussion}
\label{sec:discussion-econobio}
Our results indicate that dynamic models embedding eco-evolutionary processes, commonly used to characterise the dynamics of biological communities \cite{Bunin2017,Scheffer2006a,Case1990,Tilman1994b,eigen1988molecular}, can also appropriately characterise the dynamics of economic activities (\cref{fig:fits}). When combined with state-of-the-art computational methods and 59 years of economic data, the models provide a framework to assess the effect of eco-evolutionary processes on the long-term development of national economies.
Over the 77 countries investigated, the null model -- embedding replication and self-limitation processes only -- was rejected 45 times in favour of alternative eco-evolutionary models (\cref{fig:aggregate_countries}\textbf{B}), and each of the alternative models was supported against the null model in some countries. Hence, eco-evolutionary models that couple the growth of economic activities better characterise their dynamics. This result suggests that economic change is influenced the interdependency between economic activities. The models capturing positive interactions between economic activities ($\modalphap$) and spatial dispersal ($\moddelta$) were overall given more strength-of-evidence, indicating that these processes are central in the development of economic systems. Overall, our findings suggest that biologically inspired dynamic models are well suited to characterise economic dynamics. 

The statistical support obtained for the model with positive interactions between economic activities and the associated parameter value estimates highlights that economic activities sustain the development of other economic activities, supporting the hypothesis that economic diversity promotes economic growth \cite{Saviotti2020}. $\modalphap$ was the most frequently supported model (\cref{fig:aggregate_countries}\textbf{B},\textbf{C}), suggesting that positive interactions may be ubiquitous in economic systems. We also found that $\modalphap$ was associated with high self-limitation coefficients (\cref{fig:r_b}\textbf{B}). This finding highlights that the growth of economic activities is driven by positive interactions rather than by their endogenous growth. 
%
%
Positive interactions may arise from a variety of sub-processes acting upon firms, including beneficial business interactions through supply chains \cite{Ozman2009,Saavedra2009a} and agglomeration externalities \cite{VanDerPanne2004}. Our results suggest that the effect of positive interactions acting upon firms scales up to the national level and results in economic activities sustaining each other through positive feedbacks. 

The dynamics of economic activities seem to be greatly affected by spatial dispersal as well (\cref{fig:aggregate_countries}\textbf{B},\textbf{D}), which may originate from the spatial diffusion of organisational routines \cite{Hodgson2004} and knowledge \cite{RogersEverettM2003DoI,Keller2004,Bahar2014a} across countries. 
International business expansions \cite{Andersen1993,Zahra2000,Luo2007}, labour mobility \cite{Boschma2008}, and the diffusion of innovation through social networks \cite{RogersEverettM2003DoI,Keller2004} are candidate socio-economic processes underlying the diffusion of organisational routines. However, we found that $\moddelta$ was less frequently supported against $\modnull$ than $\modalphap$ (\cref{fig:aggregate_countries}\textbf{C}). To explain this discrepancy, we hypothesise that transfers of knowledge and routines are blocked by barriers caused by differences in cognitive, organisational, social, institutional or geographic proximity between countries \cite{Boschma2005,Caragliu2016,Bahar2014a}. This hypothesis should be investigated in future work.
Overall, our results indicate that socio-economic processes involved in the transfer of knowledge and organisational routines have an important effect on economic change at the national level.

The strong discrepancies in model support observed across countries may indicate strong idiosyncratic processes, e.g. related to the countries' institutional context \cite{Acemoglu2005} or to historical contingencies \cite{Silverberg2005a}. 
We found that developing economies have a distinct, less predictable dynamical behaviour, as the likelihood of the best models showed a negative relationship with country GDP (\cref{fig:fits}\textbf{D} and \cref{table:gdp_loglikelihood}). In line with our findings, \cite{Cristelli2015} also reports a difference in dynamical regime between mature and developing economies, where the GDP dynamics of developing economies are less predictable than those of rich countries. These model with worse fits indicate that singular processes may have an important influence on the economic dynamics of developing countries.
%
%
%
%
Additionally, while we found a statistical advantage for $\modalphap$ and $\moddelta$ over the alternative models (\cref{fig:aggregate_countries}\textbf{B}--\textbf{D}), the support and strength-of-evidence for each model varied strongly across countries (\cref{fig:aggregate_countries}A), without associations with macroeconomic characteristics such as country GDP per capita. This may be due to the noise in the data, or to the specific assumptions applied. For instance, while we assumed an equal strength of positive interactions between pairs of economic sectors (\cref{eq:submodels3}), interaction strength may be pair specific, implying that the likelihood obtained for the positive interaction hypothesis is underestimated in countries with many economic activities. 
However, these discrepancies could also originate from the idiosyncrasy of the country histories and characteristics.
Taken together, our results demonstrate the importance of idiosyncracies specific to individual countries and economic activities in determining the nature and the role of eco-evolutionary processes in economic systems.

For the sake of parsimony, the dynamic community models proposed in \cref{eq:submodels1,eq:submodels2,eq:submodels3,eq:submodels4,eq:submodels5} rely on a set of assumptions, which could be relaxed to further explore eco-evolutionary processes acting upon economic activities. 
The models investigated here rely on the mean field assumptions that all economic activities interact with an equal strength, disperse spatially across countries at an equal rate, and transfer capital between each other at an equal rate.
However, similar to biological communities \cite{Bascompte2003}, economic entities interact through complex organised networks \cite{C.A.HidalgoB.Klinger,Bustos2012,Saavedra2009a,Schweitzer2009,Giuliani2007}. For instance, \cite{C.A.HidalgoB.Klinger} suggests that economic activities are related to one another through a network of relatedness. In future studies, this network could be integrated into $\modalphan$, $\modalphap$ and $\modmu$ by weighting the interactions and transfers by the relatedness of two activities. 
%
%
Likewise, spatial dispersal is likely to depend on proximity metrics between countries \cite{Boschma2005,Caragliu2016,Bahar2014a}. Generalised variants of the models investigated, as in \cref{eq:model_general}, will need to be considered to assess the importance of economic-activity relatedness and country proximity in determining economic dynamics. Additionally, variation in the parameter values through time may need to be considered, as the strength and directionality of interactions, spatial dispersal and economic-activity transformations may change throughout the temporal development of economic activities \cite{Pistorius1997}. 
Exploring more complex eco-evolutionary models will demand richer time series, however, in order to extract the information needed to constrain the additional parameters.

In conclusion, our results put forward a biologically inspired approach for understanding the mechanisms shaping the endogenous dynamics of economic systems. By combining an inverse modelling approach relying on alternative dynamic community models with temporal economic data, our study demonstrates that positive interactions between economic activities and spatial dispersal considerably influence the dynamics of economic activities at the national level, and may be fundamental drivers of economic change.
The quantitative paradigm used to study economic systems has mainly relied on models inspired by mechanics, which assume a world in equilibrium \cite{sornette2014physics}.
Evolutionary biology, focusing on nonlinear dynamical processes and emergence, seems to be a more appropriate paradigm to study the fundamental forces shaping the dynamics of economic systems. Biological concepts can help us build process-based models to test general hypotheses on organisational principles and suggest new ones. We call this promising new research field \emph{econobiology}.

\section{Code availability}
The code used in this manuscript is available online at \href{https://github.com/vboussange/econobiology}{https://github.com/vboussange/econobiology}.

\section{Acknowledgements}
L.P. and V.B. were supported by the Swiss National Science Foundation [grant 310030E\_205556].
\bibliographystyle{unsrt}

\bibliography{bib.bib}

\begin{thebibliography}{10}

\bibitem{Dopfer2007}
Kurt Dopfer and Jason Potts.
\newblock {\em {The General Theory of Economic Evolution}}, volume~4.
\newblock Routledge, sep 2007.

\bibitem{Ruth1996}
Matthias Ruth.
\newblock {Evolutionary economics at the crossroads of biology and physics}.
\newblock {\em Journal of Social and Evolutionary Systems}, 19(2):125--144, jan
  1996.

\bibitem{schumpeter2017theory}
Joseph~A Schumpeter.
\newblock {\em {The theory of economic development: An inquiry into profits,
  capita I, credit, interest, and the business cycle}}.
\newblock Routledge, 1912.

\bibitem{gould1972}
Stephen~Jay Gould and Niles Eldredge.
\newblock {Punctuated equilibria: an alternative to phyletic gradualism}.
\newblock {\em Models in Paleobiology}, 1972:82--115, 1972.

\bibitem{Hodgson2019}
Geoffrey~M. Hodgson.
\newblock {\em {Evolutionary Economics}}, volume~66.
\newblock Cambridge University Press, jul 2019.

\bibitem{nelson1985evolutionary}
Richard~R Nelson.
\newblock {\em {An evolutionary theory of economic change}}.
\newblock Harvard University Press, 1985.

\bibitem{Hodgson2002}
Geoffrey~M. Hodgson.
\newblock {Darwinism in economics: from analogy to ontology}.
\newblock {\em Journal of Evolutionary Economics}, 12(3):259--281, jul 2002.

\bibitem{Veblen1898}
Thorstein Veblen.
\newblock {Why is economics not an evolutionary science?}
\newblock {\em The Quarterly Journal of Economics}, 12(4):373, jul 1898.

\bibitem{Pontarp2019}
Mikael Pontarp, {\AA}ke Br{\"{a}}nnstr{\"{o}}m, and Owen~L. Petchey.
\newblock {Inferring community assembly processes from macroscopic patterns
  using dynamic eco‐evolutionary models and Approximate Bayesian Computation
  (ABC)}.
\newblock {\em Methods in Ecology and Evolution}, 10(4):450--460, apr 2019.

\bibitem{Boussange2022a}
Victor Boussange, Pau Vilimelis, and Lo{\"{i}}c Pellissier.
\newblock {Mini-batching ecological data to improve ecosystem models with
  machine learning}.
\newblock 2022.

\bibitem{Skeels2022}
A~Skeels, W~Bach, O~Hagen, W~Jetz, and L~Pellissier.
\newblock {Temperature-dependent evolutionary speed shapes the Eevolution of
  biodiversity patterns across tetrapod radiations}.
\newblock {\em Systematic Biology}, 0(0):1--16, jul 2022.

\bibitem{Vellend2010}
Mark Vellend.
\newblock {Conceptual synthesis in community ecology}.
\newblock {\em The Quarterly Review of Biology}, 85(2):183--206, jun 2010.

\bibitem{Pistorius1997}
C.W.I. Pistorius and J.M. Utterback.
\newblock {Multi-mode interaction among technologies}.
\newblock {\em Research Policy}, 26(1):67--84, mar 1997.

\bibitem{GRIME1973}
John~Philip Grime.
\newblock {Competitive Exclusion in Herbaceous Vegetation}.
\newblock {\em Nature}, 242(5396):344--347, mar 1973.

\bibitem{Wernerfelt1989}
Birger Wernerfelt.
\newblock {From critical resources to corporate strategy}.
\newblock {\em Journal of General Management}, 14(3):4--12, 1989.

\bibitem{Callaway2002}
Ragan~M. Callaway, R.~W. Brooker, Philippe Choler, Zaal Kikvidze,
  Christopher~J. Lortie, Richard Michalet, Leonardo Paolini, Francisco~I.
  Pugnaire, Beth Newingham, Erik~T. Aschehoug, Cristina Armas, David Kikodze,
  and Bradley~J. Cook.
\newblock {Positive interactions among alpine plants increase with stress}.
\newblock {\em Nature}, 417(6891):844--848, 2002.

\bibitem{Ozman2009}
M.~Ozman.
\newblock {Inter-firm networks and innovation: a survey of literature}.
\newblock {\em Economics of Innovation and New Technology}, 18(1):39--67, jan
  2009.

\bibitem{Saavedra2009a}
Serguei Saavedra, Felix Reed-tsochas, and Brian Uzzi.
\newblock {A simple model of bipartite cooperation for ecological and
  organizational networks}.
\newblock {\em Nature}, 457(7228):436--466, 2009.

\bibitem{VanDerPanne2004}
Gerben {Van Der Panne}.
\newblock {Agglomeration externalities: Marshall versus Jacobs}.
\newblock {\em Journal of Evolutionary Economics}, 14(5):593--604, 2004.

\bibitem{Caragliu2016}
Andrea Caragliu and Peter Nijkamp.
\newblock {Space and knowledge spillovers in European regions: the impact of
  different forms of proximity on spatial knowledge diffusion}.
\newblock {\em Journal of Economic Geography}, 16(3):749--774, may 2016.

\bibitem{Cohendet2018}
Patrick Cohendet, David Grandadam, Chahira Mehouachi, and Laurent Simon.
\newblock {The local, the global and the industry common: the case of the video
  game industry}.
\newblock {\em Journal of Economic Geography}, 18(5):1045--1068, sep 2018.

\bibitem{Leibold2004}
M.~A. Leibold, M.~Holyoak, N.~Mouquet, P.~Amarasekare, J.~M. Chase, M.~F.
  Hoopes, R.~D. Holt, J.~B. Shurin, R.~Law, D.~Tilman, M.~Loreau, and
  A.~Gonzalez.
\newblock {The metacommunity concept: a framework for multi-scale community
  ecology}.
\newblock {\em Ecology Letters}, 7(7):601--613, jun 2004.

\bibitem{Andersen1993}
Otto Andersen.
\newblock {On the Internationalization Process of Firms: A Critical Analysis}.
\newblock {\em Journal of International Business Studies}, 24(2):209--231, jun
  1993.

\bibitem{Zahra2000}
Shaker~A. Zahra, R.~Duane Ireland, and Michael~A. Hitt.
\newblock {International expansion by new venture firms: international
  diversity, mode of market entry, technological learning, and performance}.
\newblock {\em Academy of Management Journal}, 43(5):925--950, oct 2000.

\bibitem{Luo2007}
Yadong Luo and Rosalie~L. Tung.
\newblock {International expansion of emerging market enterprises: A
  springboard perspective}.
\newblock {\em Journal of International Business Studies}, 38(4):481--498, jul
  2007.

\bibitem{Boschma2008}
Ron Boschma, Rikard Eriksson, and Urban Lindgren.
\newblock {How does labour mobility affect the performance of plants? The
  importance of relatedness and geographical proximity}.
\newblock {\em Journal of Economic Geography}, 9(2):169--190, aug 2008.

\bibitem{RogersEverettM2003DoI}
Everett~M Rogers.
\newblock {\em {Diffusion of Innovations}}.
\newblock Free Press, Riverside, 5th ed edition, 2003.

\bibitem{Keller2004}
Wolfgang Keller.
\newblock {International Technology Diffusion}.
\newblock {\em Journal of Economic Literature}, 42(3):752--782, aug 2004.

\bibitem{Bahar2014a}
Dany Bahar, Ricardo Hausmann, and Cesar~A. Hidalgo.
\newblock {Neighbors and the evolution of the comparative advantage of nations:
  Evidence of international knowledge diffusion?}
\newblock {\em Journal of International Economics}, 92(1):111--123, jan 2014.

\bibitem{Bell2017}
Graham Bell.
\newblock {Evolutionary Rescue}.
\newblock {\em Annual Review of Ecology, Evolution, and Systematics},
  48(1):605--627, nov 2017.

\bibitem{Cordes2006}
Christian Cordes.
\newblock {Darwinism in economics: from analogy to continuity}.
\newblock {\em Journal of Evolutionary Economics}, 16(5):529--541, oct 2006.

\bibitem{Freeman2002}
Christopher Freeman.
\newblock {\em {As time goes by : from the industrial revolutions to the
  information revolution}}.
\newblock Oxford University Press, Oxford, 2002.

\bibitem{Hodgson2004}
Geoffrey~M. Hodgson and Thorbjorn Knudsen.
\newblock {The firm as an interactor: firms as vehicles for habits and
  routines}.
\newblock {\em Journal of Evolutionary Economics}, 14(3):281--307, jul 2004.

\bibitem{Aldrich2008}
Howard~E. Aldrich, Geoffrey~M. Hodgson, David~L. Hull, Thorbj{\o}rn Knudsen,
  Joel Mokyr, and Viktor~J. Vanberg.
\newblock {In defence of generalized Darwinism}.
\newblock {\em Journal of Evolutionary Economics}, 18(5):577--596, 2008.

\bibitem{10.1093/cje/bet027}
Tony Lawson.
\newblock {What is this ‘school' called neoclassical economics?}
\newblock {\em Cambridge Journal of Economics}, 37(5):947--983, 2013.

\bibitem{Tacchella2018}
A.~Tacchella, D.~Mazzilli, and L.~Pietronero.
\newblock {A dynamical systems approach to gross domestic product forecasting}.
\newblock {\em Nature Physics}, 14(8):861--865, 2018.

\bibitem{Scholl2020}
Maarten~Peter Scholl, Anisoara Calinescu, and J.~Doyne Farmer.
\newblock {How market ecology explains market malfunction}.
\newblock {\em Proceedings of the National Academy of Sciences of the United
  States of America}, 118(26):e2015574118, jun 2021.

\bibitem{Zhang2018}
Guanglu Zhang, Daniel~A. McAdams, Venkatesh Shankar, and Milad {Mohammadi
  Darani}.
\newblock {Technology Evolution Prediction Using Lotka–Volterra Equations}.
\newblock {\em Journal of Mechanical Design}, 140(6):1--9, jun 2018.

\bibitem{Modis1997}
Theodore Modis.
\newblock {Genetic re-engineering of corporations}.
\newblock {\em Technological Forecasting and Social Change}, 56(2):107--118,
  oct 1997.

\bibitem{Saavedra2014}
S.~Saavedra, R.~P. Rohr, L.~J. Gilarranz, and J.~Bascompte.
\newblock {How structurally stable are global socioeconomic systems?}
\newblock {\em Journal of The Royal Society Interface},
  11(100):20140693--20140693, aug 2014.

\bibitem{Farmer1999}
J.~D. Farmer and A.~W. Lo.
\newblock {Frontiers of finance: Evolution and efficient markets}.
\newblock {\em Proceedings of the National Academy of Sciences},
  96(18):9991--9992, aug 1999.

\bibitem{Michalakelis2011}
Christos Michalakelis, Thomas Sphicopoulos, and Dimitris Varoutas.
\newblock {Modeling competition in the telecommunications market based on
  concepts of population biology}.
\newblock {\em IEEE Transactions on Systems, Man and Cybernetics Part C:
  Applications and Reviews}, 41(2):200--210, 2011.

\bibitem{Marasco2016}
A.~Marasco, A.~Picucci, and A.~Romano.
\newblock {Market share dynamics using Lotka–Volterra models}.
\newblock {\em Technological Forecasting and Social Change}, 105:49--62, apr
  2016.

\bibitem{Gatabazi2019}
P.~Gatabazi, J.C. Mba, E.~Pindza, and C.~Labuschagne.
\newblock {Grey Lotka–Volterra models with application to cryptocurrencies
  adoption}.
\newblock {\em Chaos, Solitons \& Fractals}, 122:47--57, may 2019.

\bibitem{Cauwels56}
Peter Cauwels and Didier Sornette.
\newblock {Quis Pendit Ipsa Pretia: Facebook Valuation and Diagnostic of a
  Bubble Based on Nonlinear Demographic Dynamics}.
\newblock {\em The Journal of Portfolio Management}, 38(2):56--66, jan 2012.

\bibitem{Applegate2021}
J.~M. Applegate and Adam Lampert.
\newblock {Firm size populations modeled through competition-colonization
  dynamics}.
\newblock {\em Journal of Evolutionary Economics}, 31(1):91--116, jan 2021.

\bibitem{Suweis2015}
Samir Suweis, Joel~A Carr, Amos Maritan, Andrea Rinaldo, and Paolo D'Odorico.
\newblock {Resilience and reactivity of global food security}.
\newblock {\em Proceedings of the National Academy of Sciences},
  112(22):6902--6907, jun 2015.

\bibitem{Mealy2019}
Penny Mealy, J.~Doyne Farmer, and Alexander Teytelboym.
\newblock {Interpreting economic complexity}.
\newblock {\em Science Advances}, 5(1):eaau1705, jan 2019.

\bibitem{Hidalgo2021}
C{\'{e}}sar~A. Hidalgo.
\newblock {Economic complexity theory and applications}.
\newblock {\em Nature Reviews Physics}, 3(2):92--113, feb 2021.

\bibitem{Bunin2017}
Guy Bunin.
\newblock {Ecological communities with Lotka-Volterra dynamics}.
\newblock {\em Physical Review E}, 95(4):042414, apr 2017.

\bibitem{Scheffer2006a}
Marten Scheffer and Egbert~H van Nes.
\newblock {Self-organized similarity, the evolutionary emergence of groups of
  similar species.}
\newblock {\em Proceedings of the National Academy of Sciences of the United
  States of America}, 103(16):6230--5, apr 2006.

\bibitem{Case1990}
T.~J. Case.
\newblock {Invasion resistance arises in strongly interacting species-rich
  model competition communities.}
\newblock {\em Proceedings of the National Academy of Sciences},
  87(24):9610--9614, dec 1990.

\bibitem{Tilman1994b}
David Tilman.
\newblock {Competition and Biodiversity in Spatially Structured Habitats}.
\newblock {\em Ecology}, 75(1):2--16, jan 1994.

\bibitem{eigen1988molecular}
Manfred Eigen, John McCaskill, and Peter Schuster.
\newblock {Molecular quasi-species}.
\newblock {\em The Journal of Physical Chemistry}, 92(24):6881--6891, 1988.

\bibitem{Schartau2017}
Markus Schartau, Philip Wallhead, John Hemmings, Ulrike L{\"{o}}ptien, Iris
  Kriest, Shubham Krishna, Ben~A. Ward, Thomas Slawig, and Andreas Oschlies.
\newblock {Reviews and syntheses: parameter identification in marine planktonic
  ecosystem modelling}.
\newblock {\em Biogeosciences}, 14(6):1647--1701, mar 2017.

\bibitem{Burnham2002}
David R.~Anderson {Kenneth P. Burnham} and Model.
\newblock {\em {Model Selection and Multimodel Inference}}.
\newblock Springer New York, New York, NY, 2002.

\bibitem{Kingma2014}
Diederik~P. Kingma and Jimmy Ba.
\newblock {Adam: A Method for Stochastic Optimization}.
\newblock pages 1--15, dec 2014.

\bibitem{fletcher2013practical}
Roger Fletcher.
\newblock {\em {Practical methods of optimization}}.
\newblock John Wiley \& Sons, 2013.

\bibitem{Mangan2017}
N.~M. Mangan, J.~N. Kutz, S.~L. Brunton, and J.~L. Proctor.
\newblock {Model selection for dynamical systems via sparse regression and
  information criteria}.
\newblock {\em Proceedings of the Royal Society A: Mathematical, Physical and
  Engineering Sciences}, 473(2204):20170009, aug 2017.

\bibitem{Saviotti2020}
Pier-Paolo Saviotti, Andreas Pyka¤, and Bogang Jun.
\newblock {Diversification, structural change, and economic development}.
\newblock {\em Journal of Evolutionary Economics}, 30(5):1301--1335, nov 2020.

\bibitem{Boschma2005}
Ron Boschma.
\newblock {Proximity and Innovation: A Critical Assessment}.
\newblock {\em Regional Studies}, 39(1):61--74, feb 2005.

\bibitem{Acemoglu2005}
Daron Acemoglu, Simon Johnson, and James~A. Robinson.
\newblock {Chapter 6 Institutions as a Fundamental Cause of Long-Run Growth}.
\newblock In {\em Handbook of Economic Growth}, volume~1, pages 385--472. 2005.

\bibitem{Silverberg2005a}
Gerald Silverberg and Bart Verspagen.
\newblock {Evolutionary theorizing on economic growth}.
\newblock {\em The Evolutionary Foundations of Economics}, (August):506--539,
  2005.

\bibitem{Cristelli2015}
Matthieu Cristelli, Andrea Tacchella, and Luciano Pietronero.
\newblock {The heterogeneous dynamics of economic complexity}.
\newblock {\em PLoS ONE}, 10(2):e0117174, feb 2015.

\bibitem{Bascompte2003}
J.~Bascompte, P.~Jordano, C.~J. Melian, and J.~M. Olesen.
\newblock {The nested assembly of plant-animal mutualistic networks}.
\newblock {\em Proceedings of the National Academy of Sciences},
  100(16):9383--9387, 2003.

\bibitem{C.A.HidalgoB.Klinger}
C.~A. Hidalgo, B.~Klinger, A.-L. Barabasi, and R.~Hausmann.
\newblock {The Product Space Conditions the Development of Nations}.
\newblock {\em Science}, 317(5837):482--487, jul 2007.

\bibitem{Bustos2012}
Sebasti{\'{a}}n Bustos, Charles Gomez, Ricardo Hausmann, and C{\'{e}}sar~A.
  Hidalgo.
\newblock {The Dynamics of Nestedness Predicts the Evolution of Industrial
  Ecosystems}.
\newblock {\em PLoS ONE}, 7(11):e49393, nov 2012.

\bibitem{Schweitzer2009}
Frank Schweitzer, Giorgio Fagiolo, Didier Sornette, and Douglas~R. White.
\newblock {Economic Networks : The New Challenges}.
\newblock {\em Science}, (July):422--426, 2009.

\bibitem{Giuliani2007}
Elisa Giuliani.
\newblock {The selective nature of knowledge networks in clusters: evidence
  from the wine industry}.
\newblock {\em Journal of Economic Geography}, 7(2):139--168, mar 2007.

\bibitem{sornette2014physics}
Didier Sornette.
\newblock {Physics and financial economics (1776--2014): puzzles, Ising and
  agent-based models}.
\newblock {\em Reports on Progress in Physics}, 77(6):62001, 2014.

\end{thebibliography}

\setcounter{equation}{0}
\setcounter{figure}{0}
\setcounter{table}{0}
\setcounter{page}{1}
\setcounter{section}{0}

\makeatletter

\begin{appendices}
    \renewcommand{\thetable}{S\arabic{table}}
    \renewcommand{\theequation}{S\arabic{equation}}
    \renewcommand{\thefigure}{S\arabic{figure}}
    \renewcommand{\thesection}{S\arabic{section}}
    \clearpage

\section{Supplementary Figures}
\FloatBarrier

\begin{figure}
  \center
  \includegraphics{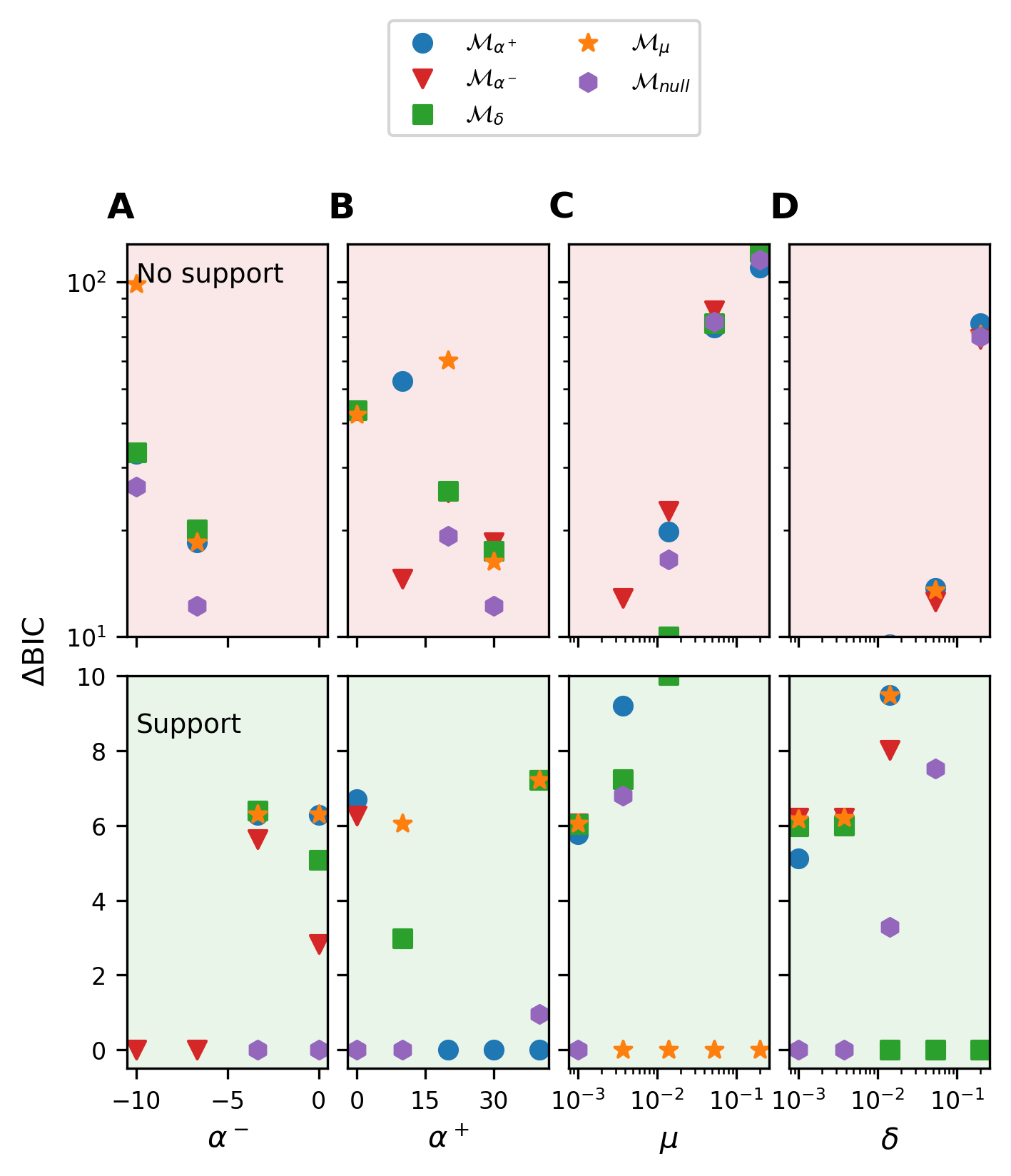}
  \caption{\small \textbf{Validation of the model selection procedure}. The figure is analogous to \cref{fig:synthetic_test_all_AIC} but with $\sigma = 0.3$. 
   }\label{figSI:synthetic_test_all_AIC}
\end{figure}

\FloatBarrier
\begin{figure}
  \center
  \includegraphics{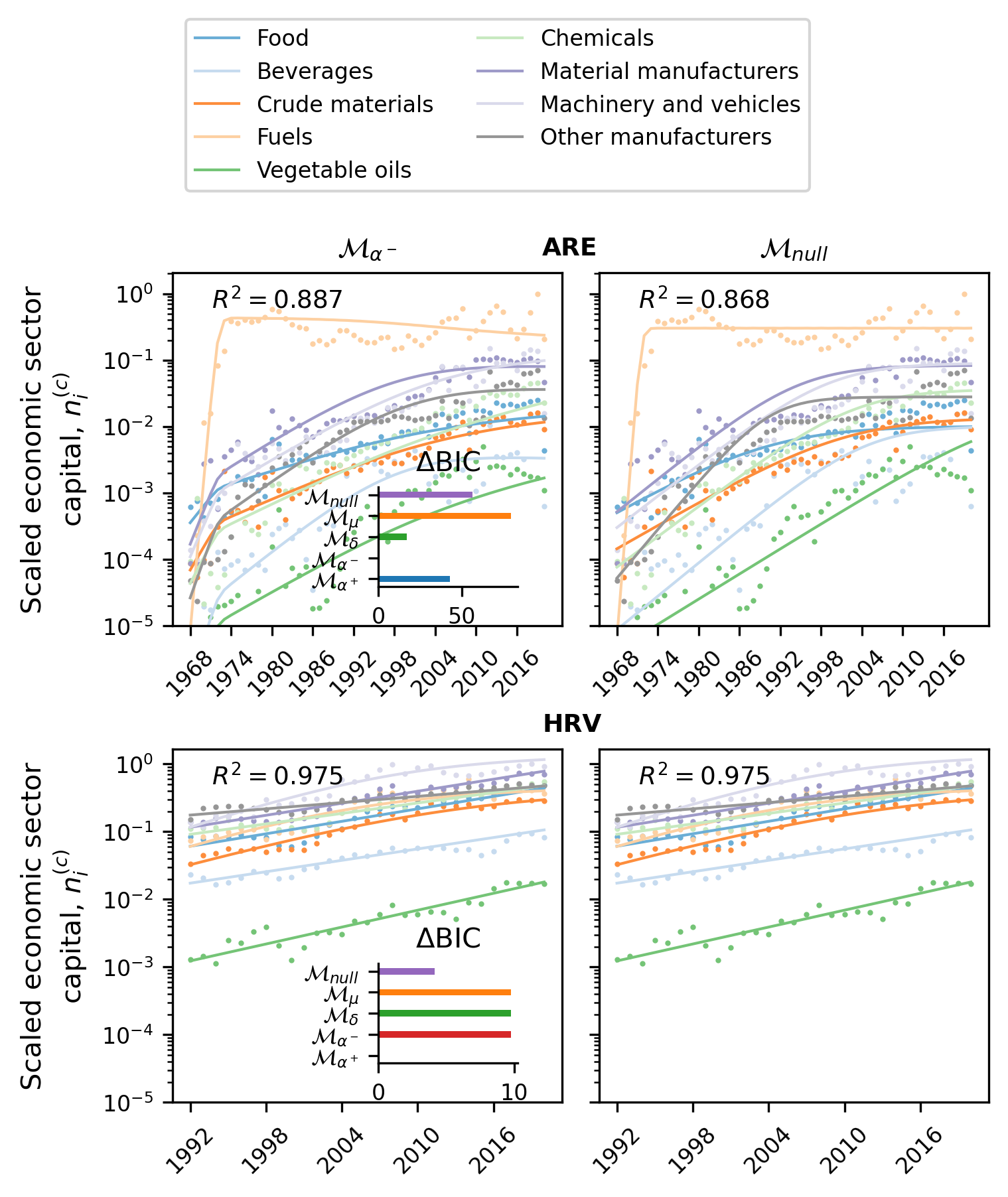}
  \caption{\small \textbf{Fits of model $\modalphan$ for United Arab Emirates (ARE) and Croatia (HRV), where $\modalphan$ is given support against $\modnull$}. 
   }\label{figSI:fit_alphan}
\end{figure}

\FloatBarrier
\begin{figure}
  \center
  \includegraphics{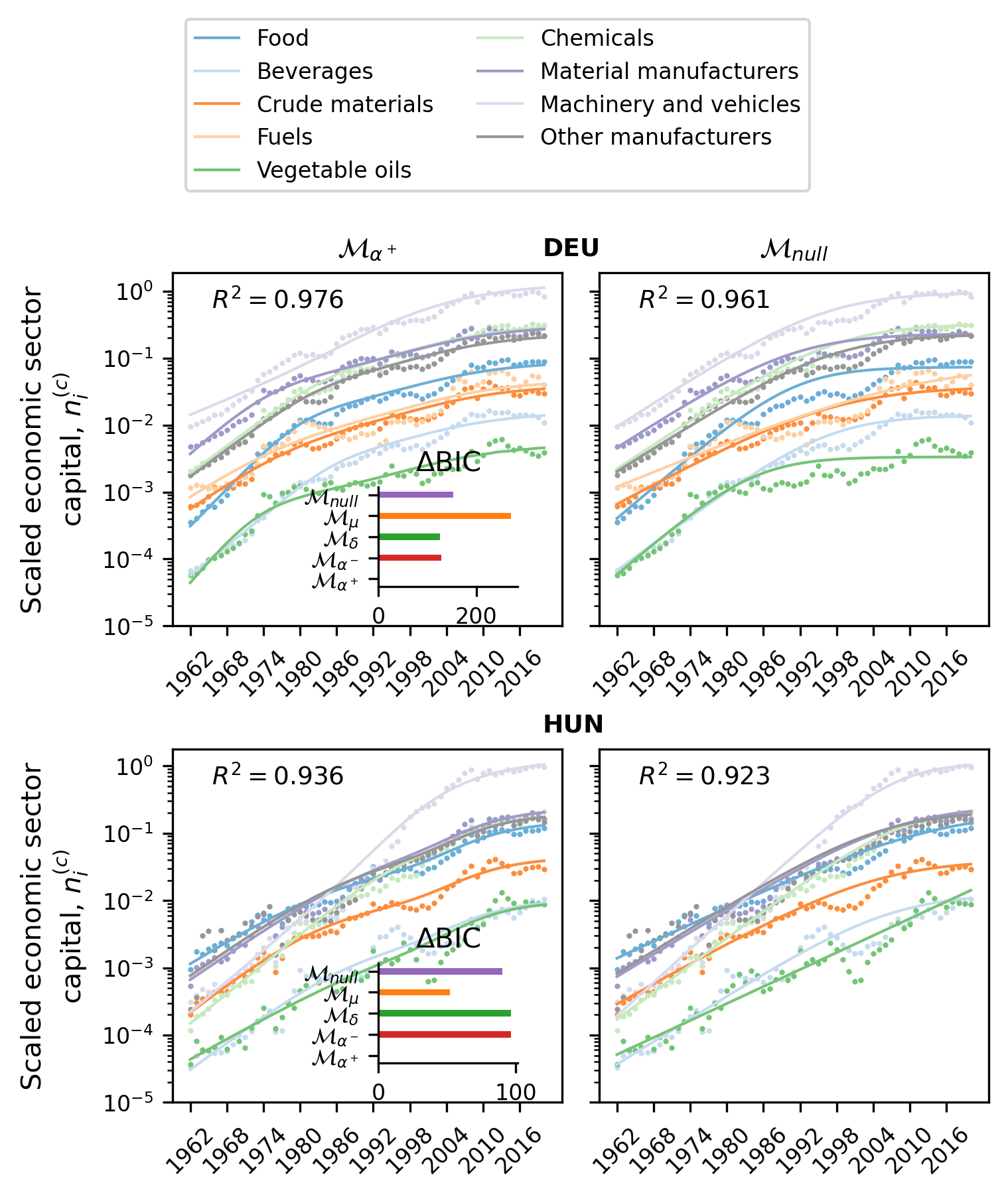}
  \caption{\small \textbf{Fits of model $\modalphap$ for Germany (DEU) and Hungaria (HUN), where $\modalphap$ is given support against $\modnull$}. 
   }\label{figSI:fit_alphap}
\end{figure}

\FloatBarrier
\begin{figure}
  \center
  \includegraphics{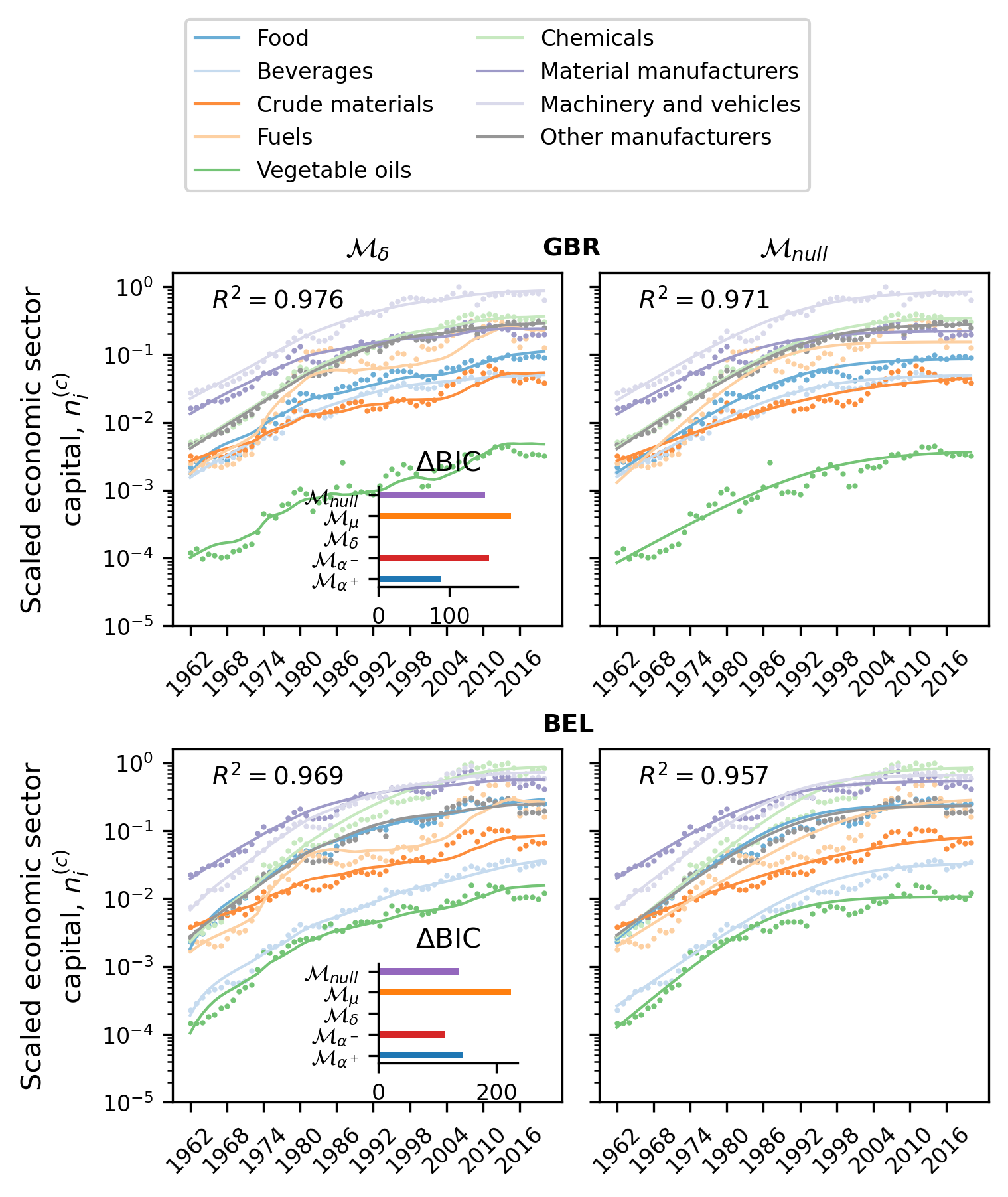}
  \caption{\small \textbf{Fits of model $\moddelta$ for Great Britain (GBR) and Belgium (BEL), where $\moddelta$ is given support against $\modnull$}. 
   }\label{figSI:fit_delta}
\end{figure}

\FloatBarrier
\begin{figure}
  \center
  \includegraphics{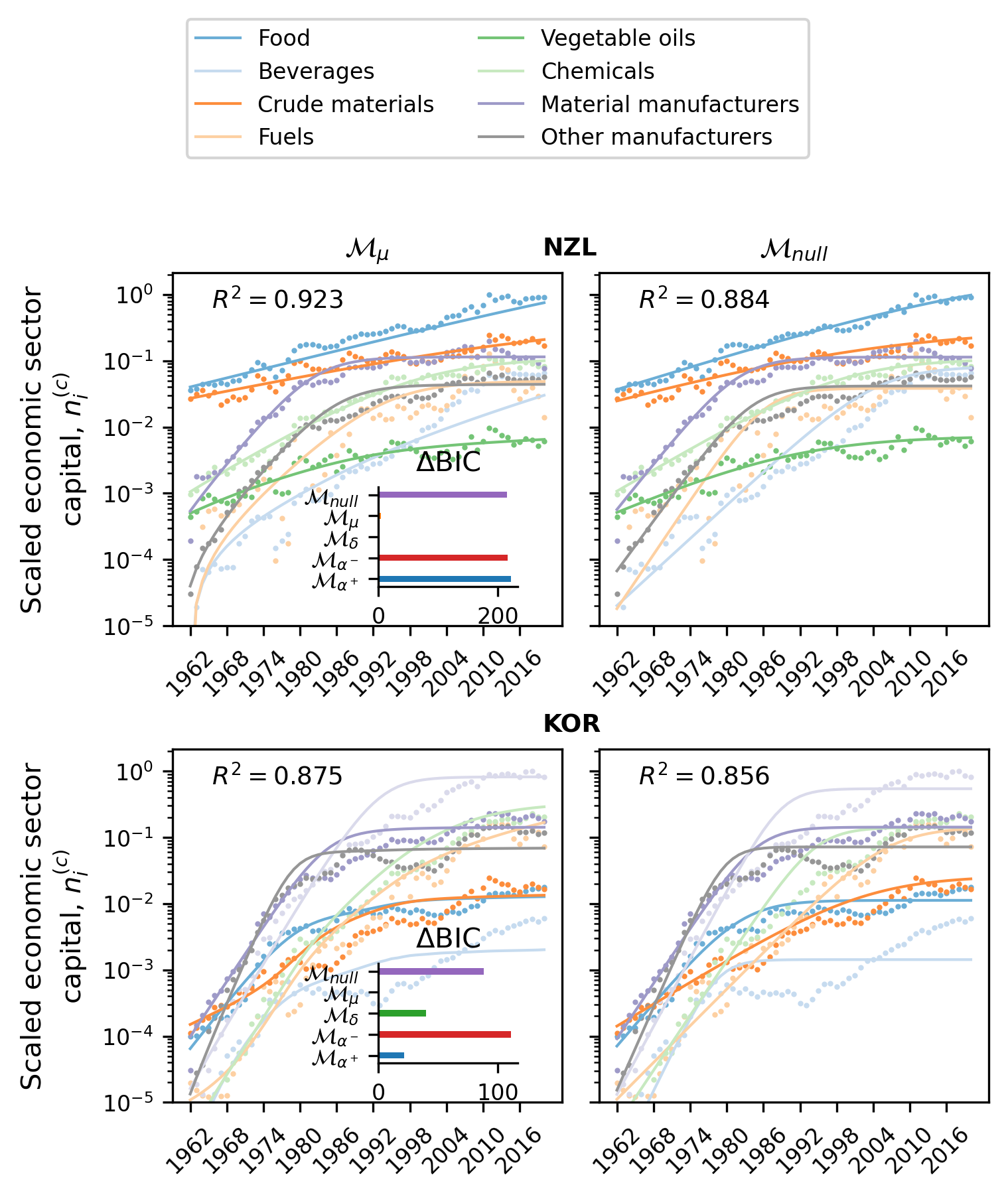}
  \caption{\small \textbf{Fits of model $\modmu$ for New Zealand (NZL) and Korea (KOR), where $\modmu$ is given support against $\modnull$}. 
   }\label{figSI:fit_mu}
\end{figure}
\FloatBarrier





\section{Supplementary Tables}

\begin{table}[ht]
    \centering
    \begin{scriptsize}
\pgfplotstabletypeset[
columns={Country Code, Country Name, Country Code, Country Name},
every head row/.style={
                        before row=\hline,
                        after row=\hline},
every last row/.style={after row=\hline},
every row/.style={after row=\hline},
display columns/0/.style={select equal part entry of={0}{2},string type, column type={|c}},
display columns/1/.style={select equal part entry of={0}{2},string type,column type={|c}},
display columns/2/.style={select equal part entry of={1}{2},string type,column type={|c}},
display columns/3/.style={select equal part entry of={1}{2},string type,column type={|c|}},
]{\datatable}
\end{scriptsize}

    \caption{\textbf{ISO 3166-1 country codes for each of the 96 countries investigated}.}
    \label{tab:country_codes}
\end{table}
\newpage

\setlength\LTcapwidth{\linewidth}
\begin{longtable}[ht]{ccccccc}
    \hline
        Country & Model & Loss & Rel. std. loss & $R^2$ & Log-likelihood & $\Delta \text{BIC}$\\
    \hline
    \endhead
    \endfoot
    \hline
    \caption{\textbf{Results of the numerical simulations}. Numerical values for each model and each country correspond to the optimisation run with the minimum loss value. "Rel. std. loss" corresponds to the standard deviation of the loss values obtained for the five independent runs, divided by the mean loss value. A low "Rel. std. loss" value indicates that the optimisation runs converged towards the same maximum likelihood.}
    \label{tab:results}
    \endlastfoot
    ARE & $\mathcal{M}_{\alpha^+}$ & 1.2527837 & 0.0443378 & 0.8985771 & 2194.9947804 & -331.4965062\\
ARE & $\mathcal{M}_{\alpha^-}$ & 1.1323544 & 0.0001987 & 0.9070505 & 2209.3385519 & -374.6811568\\
ARE & $\mathcal{M}_{\delta}$ & 1.1547997 & 0.0018924 & 0.9037398 & 2204.4213038 & -357.3571273\\
ARE & $\mathcal{M}_{\mu}$ & 1.3237895 & 0.0096762 & 0.8908284 & 2175.1338748 & -295.0532417\\
ARE & $\mathcal{M}_{null}$ & 1.3210473 & 0.0185026 & 0.8945233 & 2184.4961044 & -318.3014521\\
ARG & $\mathcal{M}_{\alpha^+}$ & 0.5362716 & 0.0000000 & 0.9126885 & 873.2783542 & -479.0901860\\
ARG & $\mathcal{M}_{\alpha^-}$ & 0.5260177 & 0.0000000 & 0.9148312 & 877.0708569 & -486.6684522\\
ARG & $\mathcal{M}_{\delta}$ & 0.5347459 & 0.0882225 & 0.9127092 & 873.5096285 & -479.1622984\\
ARG & $\mathcal{M}_{\mu}$ & 0.5362716 & 0.0000000 & 0.9126884 & 873.2790114 & -479.0898098\\
ARG & $\mathcal{M}_{null}$ & 0.5362716 & 0.0000000 & 0.9126885 & 873.2797678 & -484.8104788\\
AUS & $\mathcal{M}_{\alpha^+}$ & 0.2493686 & 0.0259488 & 0.9701803 & 1746.6372258 & -1152.0686387\\
AUS & $\mathcal{M}_{\alpha^-}$ & 0.2639190 & 0.0000105 & 0.9687099 & 1734.9134560 & -1128.5804718\\
AUS & $\mathcal{M}_{\delta}$ & 0.3009456 & 0.0013734 & 0.9620972 & 1683.0297057 & -1035.0200199\\
AUS & $\mathcal{M}_{\mu}$ & 0.2639179 & 0.0000000 & 0.9687119 & 1734.9252830 & -1128.6123545\\
AUS & $\mathcal{M}_{null}$ & 0.2639178 & 0.0000001 & 0.9687114 & 1734.9245624 & -1134.7946683\\
AUT & $\mathcal{M}_{\alpha^+}$ & 0.2068390 & 0.0001061 & 0.9818722 & 2225.2530696 & -1414.8850719\\
AUT & $\mathcal{M}_{\alpha^-}$ & 0.1927354 & 0.0000609 & 0.9831719 & 2245.3364190 & -1455.7289956\\
AUT & $\mathcal{M}_{\delta}$ & 0.1858071 & 0.0000049 & 0.9836237 & 2253.3349649 & -1470.6709565\\
AUT & $\mathcal{M}_{\mu}$ & 0.1967636 & 0.0000009 & 0.9827050 & 2238.1320372 & -1440.7057723\\
AUT & $\mathcal{M}_{null}$ & 0.2068390 & 0.0000001 & 0.9818725 & 2225.2566207 & -1421.2047025\\
BEL & $\mathcal{M}_{\alpha^+}$ & 0.1723450 & 0.0000000 & 0.9732590 & 1784.0358726 & -1495.7959922\\
BEL & $\mathcal{M}_{\alpha^-}$ & 0.1648341 & 0.0000001 & 0.9746902 & 1798.8747676 & -1525.9943707\\
BEL & $\mathcal{M}_{\delta}$ & 0.1324942 & 0.2790846 & 0.9793798 & 1854.1829608 & -1638.4943651\\
BEL & $\mathcal{M}_{\mu}$ & 0.1968834 & 0.0000000 & 0.9689876 & 1743.5335080 & -1414.4390435\\
BEL & $\mathcal{M}_{null}$ & 0.1723450 & 0.0000000 & 0.9732588 & 1784.0363416 & -1502.0995394\\
BGR & $\mathcal{M}_{\alpha^+}$ & 0.7781416 & 0.0479922 & 0.8288290 & 1764.4641596 & -649.6618052\\
BGR & $\mathcal{M}_{\alpha^-}$ & 0.8838256 & 0.0000000 & 0.8083396 & 1732.3309835 & -587.5907358\\
BGR & $\mathcal{M}_{\delta}$ & 0.8266382 & 0.0009712 & 0.8200675 & 1750.0324119 & -622.2562137\\
BGR & $\mathcal{M}_{\mu}$ & 0.8597119 & 0.0000000 & 0.8124862 & 1737.9191192 & -599.5986287\\
BGR & $\mathcal{M}_{null}$ & 0.8838256 & 0.0000922 & 0.8083396 & 1732.3319709 & -593.8988785\\
BHR & $\mathcal{M}_{\alpha^+}$ & 1.6567106 & 0.0171776 & 0.8702070 & 1089.2193585 & 55.6414398\\
BHR & $\mathcal{M}_{\alpha^-}$ & 1.5961899 & 0.0523684 & 0.8790107 & 1092.1370550 & 33.5162361\\
BHR & $\mathcal{M}_{\delta}$ & 1.6725460 & 0.0380687 & 0.8704729 & 1089.9680617 & 54.9955669\\
BHR & $\mathcal{M}_{\mu}$ & 1.6588562 & 0.0042337 & 0.8710776 & 1090.6511513 & 53.5213442\\
BHR & $\mathcal{M}_{null}$ & 1.6091897 & 0.0288135 & 0.8782411 & 1100.0620639 & 29.7611037\\
BLR & $\mathcal{M}_{\alpha^+}$ & 0.2314568 & 0.0000000 & 0.7546141 & 302.7120092 & -265.0157361\\
BLR & $\mathcal{M}_{\alpha^-}$ & 0.2310920 & 0.0000000 & 0.7559656 & 303.1611263 & -265.8441406\\
BLR & $\mathcal{M}_{\delta}$ & 0.2312815 & 0.0000001 & 0.7550474 & 302.8040002 & -265.2808168\\
BLR & $\mathcal{M}_{\mu}$ & 0.2314569 & 0.0000000 & 0.7546081 & 302.7103969 & -265.0120688\\
BLR & $\mathcal{M}_{null}$ & 0.2314568 & 0.0000000 & 0.7546151 & 302.7104437 & -270.0269676\\
BOL & $\mathcal{M}_{\alpha^+}$ & 0.3539741 & 0.5305266 & 0.9268044 & 146.3653765 & -107.0340735\\
BOL & $\mathcal{M}_{\alpha^-}$ & 0.4496981 & 0.0000247 & 0.9048095 & 135.9258779 & -87.5909551\\
BOL & $\mathcal{M}_{\delta}$ & 0.4209790 & 0.0000005 & 0.9104235 & 138.0751496 & -92.0891975\\
BOL & $\mathcal{M}_{\mu}$ & 0.4205466 & 0.0000066 & 0.9104306 & 138.0818576 & -92.0950317\\
BOL & $\mathcal{M}_{null}$ & 0.4496985 & 0.0000003 & 0.9048066 & 135.9237089 & -91.8927180\\
BRA & $\mathcal{M}_{\alpha^+}$ & 0.3172497 & 0.0000001 & 0.9440110 & 488.0802898 & -388.2378022\\
BRA & $\mathcal{M}_{\alpha^-}$ & 0.3249407 & 0.0000000 & 0.9424996 & 485.4392675 & -383.3634446\\
BRA & $\mathcal{M}_{\delta}$ & 0.3249407 & 0.0058828 & 0.9424995 & 485.4379807 & -383.3630892\\
BRA & $\mathcal{M}_{\mu}$ & 0.3206030 & 0.0000000 & 0.9433234 & 486.8511429 & -386.0040451\\
BRA & $\mathcal{M}_{null}$ & 0.3249407 & 0.0000000 & 0.9424990 & 485.4350185 & -388.5710346\\
CAN & $\mathcal{M}_{\alpha^+}$ & 0.1420790 & 0.0000000 & 0.9813800 & 1740.6230285 & -1592.8111504\\
CAN & $\mathcal{M}_{\alpha^-}$ & 0.1348187 & 0.0000000 & 0.9826218 & 1759.3484402 & -1630.7054690\\
CAN & $\mathcal{M}_{\delta}$ & 0.1345303 & 0.0002367 & 0.9823122 & 1755.3549921 & -1621.0102338\\
CAN & $\mathcal{M}_{\mu}$ & 0.1420790 & 0.0000000 & 0.9813802 & 1740.6331982 & -1592.8186471\\
CAN & $\mathcal{M}_{null}$ & 0.1420790 & 0.0000000 & 0.9813800 & 1740.6275036 & -1599.1212257\\
CHE & $\mathcal{M}_{\alpha^+}$ & 0.2379502 & 0.0002883 & 0.9792924 & 2188.2874195 & -1290.4488236\\
CHE & $\mathcal{M}_{\alpha^-}$ & 0.2381246 & 0.0725138 & 0.9792943 & 2188.2300435 & -1290.5016663\\
CHE & $\mathcal{M}_{\delta}$ & 0.2686683 & 0.0343412 & 0.9764831 & 2154.3432229 & -1220.6077504\\
CHE & $\mathcal{M}_{\mu}$ & 0.2331059 & 0.0096259 & 0.9796038 & 2192.4685594 & -1298.7697313\\
CHE & $\mathcal{M}_{null}$ & 0.2381246 & 0.1101355 & 0.9792943 & 2188.2241859 & -1296.8080908\\
CHL & $\mathcal{M}_{\alpha^+}$ & 0.5874845 & 0.0004616 & 0.9580845 & 1049.6437073 & -446.2645519\\
CHL & $\mathcal{M}_{\alpha^-}$ & 0.6179149 & 0.0000022 & 0.9565743 & 1043.9289736 & -435.4684853\\
CHL & $\mathcal{M}_{\delta}$ & 0.6179150 & 0.0000543 & 0.9565743 & 1043.9278574 & -435.4687821\\
CHL & $\mathcal{M}_{\mu}$ & 0.6179149 & 0.0000583 & 0.9565745 & 1043.9242279 & -435.4695969\\
CHL & $\mathcal{M}_{null}$ & 0.6179149 & 0.0000344 & 0.9565743 & 1043.9283281 & -441.1885246\\
COL & $\mathcal{M}_{\alpha^+}$ & 0.4490345 & 0.0000000 & 0.7782291 & 295.7157945 & -221.5722013\\
COL & $\mathcal{M}_{\alpha^-}$ & 0.4690859 & 0.0000000 & 0.7690281 & 293.0967240 & -216.6127614\\
COL & $\mathcal{M}_{\delta}$ & 0.4690859 & 0.0000000 & 0.7690293 & 293.0990791 & -216.6134084\\
COL & $\mathcal{M}_{\mu}$ & 0.4690859 & 0.0000000 & 0.7690286 & 293.0969254 & -216.6130390\\
COL & $\mathcal{M}_{null}$ & 0.4690859 & 0.0000000 & 0.7690282 & 293.0970732 & -221.4168595\\
CRI & $\mathcal{M}_{\alpha^+}$ & 2.1618680 & 0.1028749 & 0.8450607 & 1399.4320760 & -59.4886382\\
CRI & $\mathcal{M}_{\alpha^-}$ & 3.0070923 & 0.0161348 & 0.7967607 & 1355.2447368 & 38.1976757\\
CRI & $\mathcal{M}_{\delta}$ & 3.2498039 & 0.0035771 & 0.7495929 & 1312.1428722 & 113.3310469\\
CRI & $\mathcal{M}_{\mu}$ & 2.6857576 & 0.0535064 & 0.7957142 & 1349.6977712 & 40.0465823\\
CRI & $\mathcal{M}_{null}$ & 3.0070923 & 0.0000000 & 0.7967602 & 1355.2396475 & 32.3123594\\
CUB & $\mathcal{M}_{\alpha^+}$ & 0.9908768 & 0.0000001 & 0.7215276 & 478.2930704 & -171.2727897\\
CUB & $\mathcal{M}_{\alpha^-}$ & 0.8902225 & 0.0000001 & 0.7426569 & 485.8364570 & -185.7130367\\
CUB & $\mathcal{M}_{\delta}$ & 0.9908767 & 0.1727308 & 0.7215307 & 478.2948060 & -171.2747793\\
CUB & $\mathcal{M}_{\mu}$ & 0.9585458 & 0.0000020 & 0.7316088 & 481.3950009 & -178.0205603\\
CUB & $\mathcal{M}_{null}$ & 0.9908762 & 0.0000003 & 0.7215403 & 478.3330403 & -176.4906120\\
CZE & $\mathcal{M}_{\alpha^+}$ & 0.0580083 & 0.0054248 & 0.9869966 & 1018.8726966 & -778.2751708\\
CZE & $\mathcal{M}_{\alpha^-}$ & 0.0904964 & 0.0069555 & 0.9776120 & 947.2785419 & -636.4698424\\
CZE & $\mathcal{M}_{\delta}$ & 0.0898238 & 0.0017040 & 0.9775473 & 945.9773418 & -635.7157255\\
CZE & $\mathcal{M}_{\mu}$ & 0.0886011 & 0.0000000 & 0.9790624 & 955.8358104 & -653.9506041\\
CZE & $\mathcal{M}_{null}$ & 0.0904964 & 0.0000000 & 0.9776120 & 947.2785787 & -642.0340782\\
DEU & $\mathcal{M}_{\alpha^+}$ & 0.1477448 & 0.2011817 & 0.9809597 & 2336.7255710 & -1555.4873746\\
DEU & $\mathcal{M}_{\alpha^-}$ & 0.1947722 & 0.0000414 & 0.9759068 & 2272.0947772 & -1426.2671483\\
DEU & $\mathcal{M}_{\delta}$ & 0.1853032 & 0.0018003 & 0.9760610 & 2275.9987490 & -1429.7922153\\
DEU & $\mathcal{M}_{\mu}$ & 0.2447523 & 0.1317466 & 0.9688203 & 2202.4488643 & -1284.7143596\\
DEU & $\mathcal{M}_{null}$ & 0.2020155 & 0.0000000 & 0.9745613 & 2257.6743661 & -1402.7410593\\
DNK & $\mathcal{M}_{\alpha^+}$ & 0.1199929 & 0.0000000 & 0.9750129 & 1737.6862271 & -1681.0788644\\
DNK & $\mathcal{M}_{\alpha^-}$ & 0.1114577 & 0.0012059 & 0.9770632 & 1761.0406979 & -1728.0820104\\
DNK & $\mathcal{M}_{\delta}$ & 0.1067551 & 0.0075976 & 0.9778510 & 1770.9719428 & -1747.2704361\\
DNK & $\mathcal{M}_{\mu}$ & 0.1199929 & 0.0001201 & 0.9750130 & 1737.6879470 & -1681.0794476\\
DNK & $\mathcal{M}_{null}$ & 0.1199929 & 0.0000000 & 0.9750128 & 1737.6829698 & -1687.3839457\\
DOM & $\mathcal{M}_{\alpha^+}$ & 0.6961046 & 0.0003898 & 0.9155814 & 373.7708776 & -261.0889773\\
DOM & $\mathcal{M}_{\alpha^-}$ & 0.7001650 & 0.0000000 & 0.9152206 & 373.2591147 & -260.3086751\\
DOM & $\mathcal{M}_{\delta}$ & 0.7001650 & 0.0000001 & 0.9152197 & 373.2484394 & -260.3065475\\
DOM & $\mathcal{M}_{\mu}$ & 0.7001650 & 0.0000000 & 0.9152199 & 373.2546911 & -260.3071059\\
DOM & $\mathcal{M}_{null}$ & 0.7001650 & 0.0000000 & 0.9152205 & 373.2634442 & -265.5179004\\
DZA & $\mathcal{M}_{\alpha^+}$ & 1.2993726 & 0.0001052 & 0.9607808 & 442.7559336 & -90.4793137\\
DZA & $\mathcal{M}_{\alpha^-}$ & 1.2997836 & 0.0000901 & 0.9607793 & 442.6671504 & -90.4747253\\
DZA & $\mathcal{M}_{\delta}$ & 1.2996518 & 0.0001250 & 0.9607617 & 442.6369237 & -90.4198633\\
DZA & $\mathcal{M}_{\mu}$ & 1.2996155 & 0.0000682 & 0.9607943 & 442.7400231 & -90.5214100\\
DZA & $\mathcal{M}_{null}$ & 1.2996035 & 0.0000559 & 0.9607710 & 442.4031491 & -95.2526666\\
ECU & $\mathcal{M}_{\alpha^+}$ & 1.5337628 & 0.0523325 & 0.9297865 & 401.2774868 & -67.1390744\\
ECU & $\mathcal{M}_{\alpha^-}$ & 1.6887306 & 0.0000045 & 0.9209043 & 384.0521122 & -47.8421360\\
ECU & $\mathcal{M}_{\delta}$ & 1.5963688 & 0.0266030 & 0.9251918 & 390.1548236 & -56.8705370\\
ECU & $\mathcal{M}_{\mu}$ & 1.5439082 & 0.0000015 & 0.9279190 & 392.9889600 & -62.8867317\\
ECU & $\mathcal{M}_{null}$ & 1.7791235 & 0.0279990 & 0.9178951 & 382.7367272 & -46.8807035\\
ESP & $\mathcal{M}_{\alpha^+}$ & 0.1856080 & 0.0000028 & 0.9422837 & 1975.7328569 & -1434.0717545\\
ESP & $\mathcal{M}_{\alpha^-}$ & 0.1855373 & 0.0076910 & 0.9424047 & 1976.2639184 & -1435.2236056\\
ESP & $\mathcal{M}_{\delta}$ & 0.1548412 & 0.0000287 & 0.9517866 & 2025.4954451 & -1532.8383743\\
ESP & $\mathcal{M}_{\mu}$ & 0.1856080 & 0.0000000 & 0.9422840 & 1975.7359458 & -1434.0745080\\
ESP & $\mathcal{M}_{null}$ & 0.1856081 & 0.0000000 & 0.9422835 & 1975.7284854 & -1440.3779295\\
FIN & $\mathcal{M}_{\alpha^+}$ & 0.3280622 & 0.0797648 & 0.9728173 & 2152.1380163 & -1126.4399534\\
FIN & $\mathcal{M}_{\alpha^-}$ & 0.3360629 & 0.0000399 & 0.9730371 & 2153.3108657 & -1130.8958714\\
FIN & $\mathcal{M}_{\delta}$ & 0.2788228 & 0.0981958 & 0.9767311 & 2193.0232499 & -1211.7880069\\
FIN & $\mathcal{M}_{\mu}$ & 0.3539563 & 0.0007086 & 0.9708473 & 2130.2102048 & -1088.0271204\\
FIN & $\mathcal{M}_{null}$ & 0.3797915 & 0.0000000 & 0.9689312 & 2113.4406174 & -1059.3878652\\
FRA & $\mathcal{M}_{\alpha^+}$ & 0.1373927 & 0.0299971 & 0.9764265 & 1880.1804428 & -1599.3274779\\
FRA & $\mathcal{M}_{\alpha^-}$ & 0.1386926 & 0.0000073 & 0.9769100 & 1886.0075800 & -1610.7067588\\
FRA & $\mathcal{M}_{\delta}$ & 0.1198313 & 0.0474908 & 0.9789739 & 1914.1483197 & -1662.1121244\\
FRA & $\mathcal{M}_{\mu}$ & 0.1421161 & 0.1610016 & 0.9757525 & 1872.4761689 & -1583.8529740\\
FRA & $\mathcal{M}_{null}$ & 0.1466084 & 0.0000000 & 0.9751495 & 1865.9199173 & -1576.6741035\\
GBR & $\mathcal{M}_{\alpha^+}$ & 0.1669051 & 0.1074419 & 0.9775211 & 1884.4171476 & -1484.4759333\\
GBR & $\mathcal{M}_{\alpha^-}$ & 0.1938987 & 0.0000000 & 0.9746001 & 1850.8502461 & -1417.4039255\\
GBR & $\mathcal{M}_{\delta}$ & 0.1413371 & 0.0012231 & 0.9808639 & 1931.1521839 & -1572.8654147\\
GBR & $\mathcal{M}_{\mu}$ & 0.1999852 & 0.0000000 & 0.9731472 & 1835.8110921 & -1386.8661426\\
GBR & $\mathcal{M}_{null}$ & 0.1903921 & 0.0000000 & 0.9745839 & 1850.8009895 & -1423.3628352\\
GHA & $\mathcal{M}_{\alpha^+}$ & 0.3490916 & 0.0000916 & 0.7666365 & 195.2026830 & -249.8849106\\
GHA & $\mathcal{M}_{\alpha^-}$ & 0.3531896 & 0.0000000 & 0.7657012 & 194.8203537 & -249.3969268\\
GHA & $\mathcal{M}_{\delta}$ & 0.3076765 & 0.0053948 & 0.7871236 & 200.1576962 & -261.0949413\\
GHA & $\mathcal{M}_{\mu}$ & 0.3531896 & 0.0000000 & 0.7657012 & 194.8204784 & -249.3969083\\
GHA & $\mathcal{M}_{null}$ & 0.3531896 & 0.0000000 & 0.7657012 & 194.8205084 & -254.2009194\\
GRC & $\mathcal{M}_{\alpha^+}$ & 0.4896514 & 0.0161565 & 0.8280443 & 1354.6952066 & -793.6899983\\
GRC & $\mathcal{M}_{\alpha^-}$ & 0.5290440 & 0.0000000 & 0.8177233 & 1337.4385948 & -765.2452269\\
GRC & $\mathcal{M}_{\delta}$ & 0.4874272 & 0.0273553 & 0.8276619 & 1353.0776742 & -792.6062212\\
GRC & $\mathcal{M}_{\mu}$ & 0.5490437 & 0.0000000 & 0.8073630 & 1324.7477342 & -738.2677202\\
GRC & $\mathcal{M}_{null}$ & 0.5412571 & 0.0000001 & 0.8096391 & 1327.2211290 & -750.2581772\\
GTM & $\mathcal{M}_{\alpha^+}$ & 1.3642600 & 0.0080365 & 0.9021763 & 558.2459476 & -159.0748496\\
GTM & $\mathcal{M}_{\alpha^-}$ & 1.5874596 & 0.0000000 & 0.8823431 & 535.2421777 & -114.0311039\\
GTM & $\mathcal{M}_{\delta}$ & 1.5874598 & 0.0000000 & 0.8823458 & 535.2459962 & -114.0367881\\
GTM & $\mathcal{M}_{\mu}$ & 1.6555121 & 0.0000000 & 0.8757806 & 528.0917573 & -100.7875649\\
GTM & $\mathcal{M}_{null}$ & 1.5874596 & 0.0000000 & 0.8823431 & 535.2423513 & -119.5283970\\
HKG & $\mathcal{M}_{\alpha^+}$ & 0.2509271 & 0.0014428 & 0.9787234 & 2126.7218635 & -1260.9135186\\
HKG & $\mathcal{M}_{\alpha^-}$ & 0.2445060 & 0.0000013 & 0.9794208 & 2135.6830150 & -1279.2081293\\
HKG & $\mathcal{M}_{\delta}$ & 0.2517358 & 0.1089110 & 0.9785874 & 2124.9567552 & -1257.4134379\\
HKG & $\mathcal{M}_{\mu}$ & 0.2509189 & 0.2308280 & 0.9787245 & 2126.7325395 & -1260.9402382\\
HKG & $\mathcal{M}_{null}$ & 0.2509189 & 0.0014568 & 0.9787245 & 2126.7360861 & -1267.2474677\\
HRV & $\mathcal{M}_{\alpha^+}$ & 0.0659912 & 0.0001262 & 0.9787833 & 618.7632652 & -772.8094871\\
HRV & $\mathcal{M}_{\alpha^-}$ & 0.0696456 & 0.0000000 & 0.9780032 & 613.7143315 & -763.0604576\\
HRV & $\mathcal{M}_{\delta}$ & 0.0696456 & 0.0000271 & 0.9780032 & 613.7143075 & -763.0603906\\
HRV & $\mathcal{M}_{\mu}$ & 0.0696456 & 0.0000000 & 0.9780032 & 613.7143096 & -763.0604842\\
HRV & $\mathcal{M}_{null}$ & 0.0696456 & 0.0000345 & 0.9780032 & 613.7143409 & -768.6588048\\
HUN & $\mathcal{M}_{\alpha^+}$ & 0.2514814 & 0.0000000 & 0.9582531 & 2350.3148004 & -1119.3586592\\
HUN & $\mathcal{M}_{\alpha^-}$ & 0.3220814 & 0.0000345 & 0.9491220 & 2303.4750542 & -1022.8291131\\
HUN & $\mathcal{M}_{\delta}$ & 0.3220814 & 0.0000002 & 0.9491218 & 2303.4782840 & -1022.8276039\\
HUN & $\mathcal{M}_{\mu}$ & 0.2834268 & 0.0000000 & 0.9535596 & 2324.8152044 & -1067.3653028\\
HUN & $\mathcal{M}_{null}$ & 0.3220814 & 0.0000000 & 0.9491220 & 2303.4784333 & -1029.0201300\\
IDN & $\mathcal{M}_{\alpha^+}$ & 0.5192693 & 0.0000000 & 0.9146493 & 305.0747698 & -197.7229477\\
IDN & $\mathcal{M}_{\alpha^-}$ & 0.6211057 & 0.0000000 & 0.8971687 & 292.8324976 & -174.9917751\\
IDN & $\mathcal{M}_{\delta}$ & 0.5703114 & 0.0003749 & 0.9049279 & 298.3149198 & -184.5632250\\
IDN & $\mathcal{M}_{\mu}$ & 0.6211057 & 0.0000000 & 0.8971704 & 292.8323470 & -174.9937641\\
IDN & $\mathcal{M}_{null}$ & 0.6211057 & 0.0000000 & 0.8971695 & 292.8317590 & -179.7967128\\
IRL & $\mathcal{M}_{\alpha^+}$ & 0.2088275 & 0.0143105 & 0.9772782 & 2569.1155517 & -1359.9304403\\
IRL & $\mathcal{M}_{\alpha^-}$ & 0.2073867 & 0.0000036 & 0.9774790 & 2571.5504173 & -1364.8037711\\
IRL & $\mathcal{M}_{\delta}$ & 0.2090907 & 0.0200323 & 0.9772362 & 2568.5462746 & -1358.9156430\\
IRL & $\mathcal{M}_{\mu}$ & 0.2088306 & 0.0352422 & 0.9772776 & 2569.1090642 & -1359.9153338\\
IRL & $\mathcal{M}_{null}$ & 0.2088274 & 0.0144979 & 0.9772784 & 2569.1199148 & -1366.2417945\\
ISR & $\mathcal{M}_{\alpha^+}$ & 0.4160179 & 0.0062564 & 0.9709201 & 1893.2428702 & -1010.4705652\\
ISR & $\mathcal{M}_{\alpha^-}$ & 0.4349423 & 0.0000136 & 0.9701674 & 1885.8473780 & -996.4417954\\
ISR & $\mathcal{M}_{\delta}$ & 0.4339349 & 0.0008507 & 0.9698446 & 1882.2493491 & -990.5338945\\
ISR & $\mathcal{M}_{\mu}$ & 0.4349423 & 0.0000322 & 0.9701674 & 1885.8468109 & -996.4418756\\
ISR & $\mathcal{M}_{null}$ & 0.4349423 & 0.0000111 & 0.9701674 & 1885.8485980 & -1002.7501047\\
JOR & $\mathcal{M}_{\alpha^+}$ & 1.3140447 & 0.0000227 & 0.8158654 & 859.9004778 & -213.5884682\\
JOR & $\mathcal{M}_{\alpha^-}$ & 1.3883468 & 0.0000158 & 0.8112079 & 854.4614037 & -204.7457457\\
JOR & $\mathcal{M}_{\delta}$ & 1.3140447 & 0.0010034 & 0.8158620 & 859.9018193 & -213.5819671\\
JOR & $\mathcal{M}_{\mu}$ & 1.3044454 & 0.0040242 & 0.8169297 & 860.5974454 & -215.6406373\\
JOR & $\mathcal{M}_{null}$ & 1.3140446 & 0.0000000 & 0.8158654 & 859.9040348 & -219.4577945\\
JPN & $\mathcal{M}_{\alpha^+}$ & 0.0814135 & 0.0000000 & 0.9672378 & 794.7905570 & -850.5394247\\
JPN & $\mathcal{M}_{\alpha^-}$ & 0.0683921 & 0.0113106 & 0.9730056 & 818.4672910 & -897.7889660\\
JPN & $\mathcal{M}_{\delta}$ & 0.0513784 & 0.0000323 & 0.9785380 & 846.1711542 & -953.7491905\\
JPN & $\mathcal{M}_{\mu}$ & 0.0814137 & 0.0000001 & 0.9672373 & 794.7938193 & -850.5360642\\
JPN & $\mathcal{M}_{null}$ & 0.0814135 & 0.0000002 & 0.9672371 & 794.7868478 & -856.0313369\\
KAZ & $\mathcal{M}_{\alpha^+}$ & 0.1865516 & 0.0000476 & 0.8295751 & 214.6552620 & -174.1895588\\
KAZ & $\mathcal{M}_{\alpha^-}$ & 0.2496131 & 0.0000000 & 0.7623985 & 199.8289770 & -144.2825364\\
KAZ & $\mathcal{M}_{\delta}$ & 0.2390774 & 0.0000000 & 0.7721660 & 201.6233048 & -148.0605220\\
KAZ & $\mathcal{M}_{\mu}$ & 0.2362297 & 0.0000001 & 0.7769984 & 202.2588602 & -149.9899724\\
KAZ & $\mathcal{M}_{null}$ & 0.2496130 & 0.0000001 & 0.7624011 & 199.8317347 & -148.7833403\\
KOR & $\mathcal{M}_{\alpha^+}$ & 0.6920979 & 0.0649634 & 0.9155851 & 2216.2272046 & -646.0307164\\
KOR & $\mathcal{M}_{\alpha^-}$ & 0.8387203 & 0.0000073 & 0.8985858 & 2168.9233801 & -556.4976910\\
KOR & $\mathcal{M}_{\delta}$ & 0.6985206 & 0.0000000 & 0.9123594 & 2202.6982338 & -627.7305826\\
KOR & $\mathcal{M}_{\mu}$ & 0.6614961 & 0.2464106 & 0.9192266 & 2224.9307739 & -667.5495675\\
KOR & $\mathcal{M}_{null}$ & 0.7946315 & 0.0000000 & 0.9020007 & 2176.7551440 & -579.4032059\\
LBN & $\mathcal{M}_{\alpha^+}$ & 1.3016604 & 0.0240624 & 0.8243273 & 807.8076783 & -327.5331892\\
LBN & $\mathcal{M}_{\alpha^-}$ & 1.4386036 & 0.0000001 & 0.8073042 & 782.3615142 & -282.3978581\\
LBN & $\mathcal{M}_{\delta}$ & 1.4380388 & 0.0005367 & 0.8117306 & 789.9437981 & -293.7384027\\
LBN & $\mathcal{M}_{\mu}$ & 1.4320212 & 0.0000001 & 0.8135593 & 792.1142739 & -298.5017188\\
LBN & $\mathcal{M}_{null}$ & 1.4416777 & 0.0000000 & 0.8100418 & 788.0379102 & -295.5709325\\
LBY & $\mathcal{M}_{\alpha^+}$ & 2.2780600 & 0.0000128 & 0.9267840 & 378.4283856 & -11.0745573\\
LBY & $\mathcal{M}_{\alpha^-}$ & 2.8894278 & 0.0000473 & 0.9079340 & 364.6708243 & 16.8746157\\
LBY & $\mathcal{M}_{\delta}$ & 3.5313144 & 0.0002249 & 0.8853192 & 349.5494084 & 43.6715402\\
LBY & $\mathcal{M}_{\mu}$ & 3.4722522 & 0.0000224 & 0.8868652 & 350.6939375 & 42.0157078\\
LBY & $\mathcal{M}_{null}$ & 2.7553149 & 0.0000477 & 0.9107669 & 365.7889379 & 8.2576748\\
LKA & $\mathcal{M}_{\alpha^+}$ & 1.5941187 & 0.0003071 & 0.8115942 & 336.9975384 & -92.8885083\\
LKA & $\mathcal{M}_{\alpha^-}$ & 1.6362543 & 0.0000000 & 0.8102917 & 337.0514707 & -91.6277204\\
LKA & $\mathcal{M}_{\delta}$ & 1.6322208 & 0.0022002 & 0.8084860 & 335.3889986 & -89.8940407\\
LKA & $\mathcal{M}_{\mu}$ & 1.6362543 & 0.0025755 & 0.8102917 & 337.0514702 & -91.6277204\\
LKA & $\mathcal{M}_{null}$ & 1.6362543 & 0.0000000 & 0.8102917 & 337.0514708 & -96.8372065\\
LTU & $\mathcal{M}_{\alpha^+}$ & 0.2792254 & 0.0000068 & 0.9197774 & 550.3188221 & -359.2088757\\
LTU & $\mathcal{M}_{\alpha^-}$ & 0.2792263 & 0.0011184 & 0.9197673 & 550.3111851 & -359.1746619\\
LTU & $\mathcal{M}_{\delta}$ & 0.2539895 & 0.0000002 & 0.9241161 & 555.7579934 & -374.2208641\\
LTU & $\mathcal{M}_{\mu}$ & 0.2554806 & 0.0000000 & 0.9240155 & 556.4417023 & -373.8632704\\
LTU & $\mathcal{M}_{null}$ & 0.2792264 & 0.0000000 & 0.9197682 & 550.3064726 & -364.7762372\\
LUX & $\mathcal{M}_{\alpha^+}$ & 0.1687301 & 0.0027144 & 0.9756622 & 606.0315078 & -387.5202145\\
LUX & $\mathcal{M}_{\alpha^-}$ & 0.1556444 & 0.0423426 & 0.9776990 & 616.1184100 & -405.6120311\\
LUX & $\mathcal{M}_{\delta}$ & 0.1556444 & 0.0037217 & 0.9776974 & 616.1053905 & -405.5967384\\
LUX & $\mathcal{M}_{\mu}$ & 0.1556444 & 0.0514556 & 0.9776986 & 616.1152846 & -405.6079128\\
LUX & $\mathcal{M}_{null}$ & 0.1556444 & 0.0000000 & 0.9776985 & 616.1172377 & -410.9401680\\
LVA & $\mathcal{M}_{\alpha^+}$ & 0.2953602 & 0.0000000 & 0.9317351 & 543.8952197 & -366.6813027\\
LVA & $\mathcal{M}_{\alpha^-}$ & 0.2877696 & 0.0000698 & 0.9322366 & 544.6567294 & -368.6722851\\
LVA & $\mathcal{M}_{\delta}$ & 0.2280066 & 0.0002703 & 0.9426518 & 563.8282263 & -413.7302090\\
LVA & $\mathcal{M}_{\mu}$ & 0.2682572 & 0.0000000 & 0.9354481 & 550.4990299 & -381.7817245\\
LVA & $\mathcal{M}_{null}$ & 0.2953602 & 0.0000000 & 0.9317370 & 543.8963214 & -372.2873909\\
MAC & $\mathcal{M}_{\alpha^+}$ & 1.1294488 & 0.0004599 & 0.9315991 & 1531.4525516 & -289.1433884\\
MAC & $\mathcal{M}_{\alpha^-}$ & 0.9767320 & 0.0000041 & 0.9406784 & 1567.2754683 & -349.5263422\\
MAC & $\mathcal{M}_{\delta}$ & 1.1294497 & 0.0051249 & 0.9315974 & 1531.4577652 & -289.1325538\\
MAC & $\mathcal{M}_{\mu}$ & 1.1294485 & 0.0005140 & 0.9315978 & 1531.4587640 & -289.1354956\\
MAC & $\mathcal{M}_{null}$ & 1.1294486 & 0.0005148 & 0.9315990 & 1531.4584828 & -295.1922572\\
MAR & $\mathcal{M}_{\alpha^+}$ & 0.8537848 & 0.0234547 & 0.9349568 & 650.3604985 & -264.3866710\\
MAR & $\mathcal{M}_{\alpha^-}$ & 0.9295200 & 0.0000018 & 0.9296139 & 640.4149824 & -245.1246477\\
MAR & $\mathcal{M}_{\delta}$ & 0.9016048 & 0.0040674 & 0.9319661 & 643.9018432 & -253.4178828\\
MAR & $\mathcal{M}_{\mu}$ & 0.9173997 & 0.0000204 & 0.9310205 & 642.7052567 & -250.0499976\\
MAR & $\mathcal{M}_{null}$ & 0.9295200 & 0.0000035 & 0.9296140 & 640.4143693 & -250.6218776\\
MEX & $\mathcal{M}_{\alpha^+}$ & 0.4508260 & 0.0000993 & 0.9151172 & 1435.2035699 & -631.2102997\\
MEX & $\mathcal{M}_{\alpha^-}$ & 0.4876843 & 0.0000001 & 0.9084882 & 1420.6198791 & -603.6884578\\
MEX & $\mathcal{M}_{\delta}$ & 0.4392520 & 0.0149692 & 0.9163000 & 1437.2548352 & -636.3463466\\
MEX & $\mathcal{M}_{\mu}$ & 0.4866618 & 0.0092329 & 0.9082319 & 1420.0192358 & -602.6647468\\
MEX & $\mathcal{M}_{null}$ & 0.4876843 & 0.0000001 & 0.9084879 & 1420.6159465 & -609.5898333\\
MYS & $\mathcal{M}_{\alpha^+}$ & 0.2895046 & 0.0785445 & 0.9539404 & 1973.5958108 & -1124.8476489\\
MYS & $\mathcal{M}_{\alpha^-}$ & 0.3399726 & 0.0000000 & 0.9474011 & 1937.0418339 & -1054.3528091\\
MYS & $\mathcal{M}_{\delta}$ & 0.3399726 & 0.0000000 & 0.9474014 & 1937.0487616 & -1054.3553106\\
MYS & $\mathcal{M}_{\mu}$ & 0.3169605 & 0.0000000 & 0.9509908 & 1955.6429441 & -1091.8872434\\
MYS & $\mathcal{M}_{null}$ & 0.3399726 & 0.0000000 & 0.9474012 & 1937.0402581 & -1060.6285880\\
NGA & $\mathcal{M}_{\alpha^+}$ & 2.7990310 & 0.0000239 & 0.9395854 & 495.7796031 & 11.7048325\\
NGA & $\mathcal{M}_{\alpha^-}$ & 2.7992006 & 0.0000615 & 0.9395758 & 495.6501030 & 11.7242375\\
NGA & $\mathcal{M}_{\delta}$ & 2.6256522 & 0.0000111 & 0.9431462 & 498.4576558 & 4.2936995\\
NGA & $\mathcal{M}_{\mu}$ & 2.7268460 & 0.0000217 & 0.9411712 & 496.2771356 & 8.4598284\\
NGA & $\mathcal{M}_{null}$ & 2.7990746 & 0.0000159 & 0.9395804 & 495.7132708 & 6.9110458\\
NLD & $\mathcal{M}_{\alpha^+}$ & 0.1466304 & 0.0000001 & 0.9582107 & 1657.4935262 & -1562.8725186\\
NLD & $\mathcal{M}_{\alpha^-}$ & 0.2092032 & 0.0000000 & 0.9418981 & 1567.4351623 & -1381.9447140\\
NLD & $\mathcal{M}_{\delta}$ & 0.1340786 & 0.0351018 & 0.9609515 & 1678.9038968 & -1600.1142643\\
NLD & $\mathcal{M}_{\mu}$ & 0.2097818 & 0.0272475 & 0.9412098 & 1564.2116085 & -1375.4789279\\
NLD & $\mathcal{M}_{null}$ & 0.2097818 & 0.0000000 & 0.9412096 & 1564.2134536 & -1381.7853394\\
NOR & $\mathcal{M}_{\alpha^+}$ & 0.2399445 & 0.0130421 & 0.9780959 & 2521.5074784 & -1298.8700700\\
NOR & $\mathcal{M}_{\alpha^-}$ & 0.2390888 & 0.0016231 & 0.9781834 & 2522.6463097 & -1301.0687029\\
NOR & $\mathcal{M}_{\delta}$ & 0.2400056 & 0.0128037 & 0.9780893 & 2521.4101099 & -1298.7067950\\
NOR & $\mathcal{M}_{\mu}$ & 0.2399438 & 0.0130562 & 0.9780959 & 2521.5037000 & -1298.8707092\\
NOR & $\mathcal{M}_{null}$ & 0.2398829 & 0.0193799 & 0.9780996 & 2521.5638888 & -1305.2721046\\
NZL & $\mathcal{M}_{\alpha^+}$ & 0.9073281 & 0.0037365 & 0.9129915 & 1654.8440041 & -567.6419686\\
NZL & $\mathcal{M}_{\alpha^-}$ & 0.8772374 & 0.0000059 & 0.9140447 & 1656.6103203 & -573.5852855\\
NZL & $\mathcal{M}_{\delta}$ & 0.4805584 & 0.0203982 & 0.9447784 & 1759.6846130 & -789.5124824\\
NZL & $\mathcal{M}_{\mu}$ & 0.4917836 & 0.0120060 & 0.9442763 & 1758.4536220 & -785.0947066\\
NZL & $\mathcal{M}_{null}$ & 0.9073281 & 0.0000000 & 0.9129914 & 1654.8396788 & -573.8317817\\
OMN & $\mathcal{M}_{\alpha^+}$ & 0.7889506 & 0.0396981 & 0.9139409 & 1112.8720701 & -153.1574333\\
OMN & $\mathcal{M}_{\alpha^-}$ & 0.6700664 & 0.0860673 & 0.9239838 & 1120.4887405 & -183.5588674\\
OMN & $\mathcal{M}_{\delta}$ & 0.8933917 & 0.0678704 & 0.9031544 & 1102.8136604 & -124.2268460\\
OMN & $\mathcal{M}_{\mu}$ & 0.8791464 & 0.0000000 & 0.9020953 & 1095.4922578 & -121.5620907\\
OMN & $\mathcal{M}_{null}$ & 0.8520861 & 0.0006775 & 0.9052430 & 1100.9144521 & -135.0695316\\
PAN & $\mathcal{M}_{\alpha^+}$ & 1.1143223 & 0.0000000 & 0.8199065 & 998.7151866 & -355.1878253\\
PAN & $\mathcal{M}_{\alpha^-}$ & 1.1149436 & 0.0000000 & 0.8201169 & 998.8693793 & -355.6869349\\
PAN & $\mathcal{M}_{\delta}$ & 1.1100368 & 0.0002924 & 0.8199278 & 998.2834172 & -355.2383815\\
PAN & $\mathcal{M}_{\mu}$ & 1.1149435 & 0.0000000 & 0.8201169 & 998.8720122 & -355.6869611\\
PAN & $\mathcal{M}_{null}$ & 1.1149436 & 0.0000000 & 0.8201166 & 998.8746400 & -361.7432027\\
PER & $\mathcal{M}_{\alpha^+}$ & 0.8254032 & 0.0311618 & 0.8826033 & 573.8679392 & -274.9069780\\
PER & $\mathcal{M}_{\alpha^-}$ & 0.9361490 & 0.0000000 & 0.8671464 & 559.4016095 & -244.7267453\\
PER & $\mathcal{M}_{\delta}$ & 0.8882912 & 0.0000974 & 0.8760467 & 567.2656212 & -261.6465567\\
PER & $\mathcal{M}_{\mu}$ & 0.9055167 & 0.0000000 & 0.8725708 & 564.1687729 & -254.8983509\\
PER & $\mathcal{M}_{null}$ & 0.9361490 & 0.0000000 & 0.8671464 & 559.4031274 & -250.2239393\\
POL & $\mathcal{M}_{\alpha^+}$ & 0.2535426 & 0.0000000 & 0.9438326 & 1807.0469575 & -1001.0879697\\
POL & $\mathcal{M}_{\alpha^-}$ & 0.3163218 & 0.0000000 & 0.9326013 & 1768.7678021 & -923.2501964\\
POL & $\mathcal{M}_{\delta}$ & 0.3162856 & 0.0000403 & 0.9326143 & 1768.7760416 & -923.3322541\\
POL & $\mathcal{M}_{\mu}$ & 0.2782767 & 0.0000000 & 0.9379993 & 1785.4443202 & -958.8962784\\
POL & $\mathcal{M}_{null}$ & 0.3163218 & 0.0000000 & 0.9326025 & 1768.7577185 & -929.3142565\\
PRT & $\mathcal{M}_{\alpha^+}$ & 0.3321047 & 0.0000076 & 0.9133453 & 1726.5468270 & -1093.0105775\\
PRT & $\mathcal{M}_{\alpha^-}$ & 0.3453640 & 0.0127740 & 0.9112486 & 1719.7441003 & -1079.8850541\\
PRT & $\mathcal{M}_{\delta}$ & 0.3369898 & 0.0000004 & 0.9136215 & 1727.2952235 & -1094.7635003\\
PRT & $\mathcal{M}_{\mu}$ & 0.3417620 & 0.0000000 & 0.9114441 & 1720.4004837 & -1081.0956205\\
PRT & $\mathcal{M}_{null}$ & 0.3453640 & 0.0000000 & 0.9112488 & 1719.7422418 & -1086.1944995\\
PRY & $\mathcal{M}_{\alpha^+}$ & 0.7911077 & 0.0000299 & 0.8628434 & 618.4255027 & -283.9049804\\
PRY & $\mathcal{M}_{\alpha^-}$ & 0.8457188 & 0.0000000 & 0.8541465 & 610.2534769 & -268.9040302\\
PRY & $\mathcal{M}_{\delta}$ & 0.8457189 & 0.0000006 & 0.8541473 & 610.2555041 & -268.9052712\\
PRY & $\mathcal{M}_{\mu}$ & 0.8457188 & 0.0000003 & 0.8541469 & 610.2551282 & -268.9046708\\
PRY & $\mathcal{M}_{null}$ & 0.8457188 & 0.0000007 & 0.8541468 & 610.2536061 & -274.4016218\\
QAT & $\mathcal{M}_{\alpha^+}$ & 1.9266798 & 0.0038751 & 0.9118724 & 1618.1838222 & -63.5800421\\
QAT & $\mathcal{M}_{\alpha^-}$ & 2.0139147 & 0.0000255 & 0.9086920 & 1608.4377172 & -51.8806944\\
QAT & $\mathcal{M}_{\delta}$ & 1.9353840 & 0.0001128 & 0.9115033 & 1615.0203438 & -62.2009448\\
QAT & $\mathcal{M}_{\mu}$ & 1.9353839 & 0.0226246 & 0.9115036 & 1615.0306966 & -62.2018858\\
QAT & $\mathcal{M}_{null}$ & 1.9353843 & 0.0000104 & 0.9115025 & 1615.0092877 & -67.9971407\\
ROU & $\mathcal{M}_{\alpha^+}$ & 0.8900993 & 0.0290867 & 0.8907507 & 1747.7760599 & -447.7210801\\
ROU & $\mathcal{M}_{\alpha^-}$ & 1.0376017 & 0.0000015 & 0.8796222 & 1727.7138904 & -406.3009748\\
ROU & $\mathcal{M}_{\delta}$ & 1.0376021 & 0.0000221 & 0.8796235 & 1727.7104815 & -406.3056340\\
ROU & $\mathcal{M}_{\mu}$ & 1.0008829 & 0.0125723 & 0.8786329 & 1723.9683488 & -402.8060015\\
ROU & $\mathcal{M}_{null}$ & 1.0376017 & 0.0000001 & 0.8796230 & 1727.7051904 & -412.3607662\\
RUS & $\mathcal{M}_{\alpha^+}$ & 0.6174841 & 0.0000000 & 0.9566539 & 1070.3858708 & -349.4521333\\
RUS & $\mathcal{M}_{\alpha^-}$ & 0.8291502 & 0.0000002 & 0.9429638 & 1037.3556155 & -282.4815040\\
RUS & $\mathcal{M}_{\delta}$ & 0.8291506 & 0.0000000 & 0.9429651 & 1037.3633751 & -282.4866992\\
RUS & $\mathcal{M}_{\mu}$ & 0.8194430 & 0.0000824 & 0.9438272 & 1038.7924091 & -286.2030003\\
RUS & $\mathcal{M}_{null}$ & 0.8291502 & 0.0000000 & 0.9429641 & 1037.3532266 & -287.9799069\\
SAU & $\mathcal{M}_{\alpha^+}$ & 1.4447598 & 0.0000015 & 0.9319666 & 318.2772364 & -63.7460639\\
SAU & $\mathcal{M}_{\alpha^-}$ & 1.1467943 & 0.0000016 & 0.9491154 & 335.8594189 & -98.0178968\\
SAU & $\mathcal{M}_{\delta}$ & 1.4317991 & 0.0000000 & 0.9331044 & 319.2716499 & -65.7362927\\
SAU & $\mathcal{M}_{\mu}$ & 1.3103942 & 0.0000004 & 0.9409868 & 327.3080046 & -80.5301406\\
SAU & $\mathcal{M}_{null}$ & 1.4447565 & 0.0000027 & 0.9319483 & 318.1644064 & -68.4849842\\
SGP & $\mathcal{M}_{\alpha^+}$ & 0.4390113 & 0.0002142 & 0.9326026 & 2040.0506241 & -979.1514085\\
SGP & $\mathcal{M}_{\alpha^-}$ & 0.4389674 & 0.1024984 & 0.9330349 & 2041.7687561 & -982.6841384\\
SGP & $\mathcal{M}_{\delta}$ & 0.4381149 & 0.0988135 & 0.9327193 & 2040.8073272 & -980.1034145\\
SGP & $\mathcal{M}_{\mu}$ & 0.4390113 & 0.0001842 & 0.9326026 & 2040.0476396 & -979.1513242\\
SGP & $\mathcal{M}_{null}$ & 0.4390113 & 0.1632410 & 0.9326026 & 2040.0510546 & -985.4594657\\
SVK & $\mathcal{M}_{\alpha^+}$ & 0.0897265 & 0.2508679 & 0.9841495 & 1031.5427644 & -671.5084754\\
SVK & $\mathcal{M}_{\alpha^-}$ & 0.1076807 & 0.0000152 & 0.9797558 & 999.7833418 & -607.6491798\\
SVK & $\mathcal{M}_{\delta}$ & 0.1297093 & 0.0563428 & 0.9763511 & 979.6865381 & -567.0771140\\
SVK & $\mathcal{M}_{\mu}$ & 0.1164278 & 0.0012728 & 0.9782005 & 991.2554924 & -588.3305464\\
SVK & $\mathcal{M}_{null}$ & 0.1135388 & 0.0000000 & 0.9781674 & 991.0807478 & -593.4997133\\
SVN & $\mathcal{M}_{\alpha^+}$ & 0.1106088 & 0.0196362 & 0.9794616 & 807.4033352 & -625.2063291\\
SVN & $\mathcal{M}_{\alpha^-}$ & 0.1204472 & 0.0001807 & 0.9768129 & 791.5610114 & -592.4550281\\
SVN & $\mathcal{M}_{\delta}$ & 0.1060602 & 0.0012029 & 0.9792287 & 805.5613897 & -622.1622849\\
SVN & $\mathcal{M}_{\mu}$ & 0.1198190 & 0.0006017 & 0.9769728 & 792.4983162 & -594.3233439\\
SVN & $\mathcal{M}_{null}$ & 0.1204460 & 0.0001852 & 0.9768135 & 791.5636796 & -598.0604505\\
SWE & $\mathcal{M}_{\alpha^+}$ & 0.1794411 & 0.0000246 & 0.9827111 & 2093.2525551 & -1458.8338866\\
SWE & $\mathcal{M}_{\alpha^-}$ & 0.1679885 & 0.0011001 & 0.9840927 & 2115.9414425 & -1504.5572179\\
SWE & $\mathcal{M}_{\delta}$ & 0.1604230 & 0.0619388 & 0.9845382 & 2124.0345953 & -1520.1547048\\
SWE & $\mathcal{M}_{\mu}$ & 0.1773051 & 0.0333840 & 0.9828882 & 2095.9941251 & -1464.4867308\\
SWE & $\mathcal{M}_{null}$ & 0.1794411 & 0.0000000 & 0.9827112 & 2093.2494231 & -1465.1444556\\
THA & $\mathcal{M}_{\alpha^+}$ & 0.3214689 & 0.0000002 & 0.9515592 & 1740.3783079 & -881.8617066\\
THA & $\mathcal{M}_{\alpha^-}$ & 0.3563113 & 0.0000219 & 0.9490399 & 1729.7252912 & -860.2127021\\
THA & $\mathcal{M}_{\delta}$ & 0.3563231 & 0.0000376 & 0.9490292 & 1729.6876773 & -860.1227083\\
THA & $\mathcal{M}_{\mu}$ & 0.3563111 & 0.0000166 & 0.9490411 & 1729.7307152 & -860.2231110\\
THA & $\mathcal{M}_{null}$ & 0.3563111 & 0.0000002 & 0.9490411 & 1729.7276038 & -866.2793417\\
TKM & $\mathcal{M}_{\alpha^+}$ & 0.6120721 & 0.0002175 & 0.8484762 & 110.4929157 & -52.1194340\\
TKM & $\mathcal{M}_{\alpha^-}$ & 0.6125306 & 0.0002738 & 0.8486140 & 110.5452317 & -52.1740147\\
TKM & $\mathcal{M}_{\delta}$ & 0.6125441 & 0.0000282 & 0.8483613 & 110.4933587 & -52.0739457\\
TKM & $\mathcal{M}_{\mu}$ & 0.6125429 & 0.0000481 & 0.8483465 & 110.4568505 & -52.0681256\\
TKM & $\mathcal{M}_{null}$ & 0.6125373 & 0.0000492 & 0.8483298 & 110.4627760 & -56.1558440\\
TUN & $\mathcal{M}_{\alpha^+}$ & 0.6040435 & 0.0519402 & 0.8800402 & 970.7830183 & -525.8558531\\
TUN & $\mathcal{M}_{\alpha^-}$ & 0.6553858 & 0.0000039 & 0.8721401 & 959.5907734 & -502.5129657\\
TUN & $\mathcal{M}_{\delta}$ & 0.6413280 & 0.0134767 & 0.8740118 & 962.3288209 & -507.9103306\\
TUN & $\mathcal{M}_{\mu}$ & 0.6486814 & 0.0000013 & 0.8731389 & 960.7741688 & -505.3833771\\
TUN & $\mathcal{M}_{null}$ & 0.6553859 & 0.0000000 & 0.8721396 & 959.5919379 & -508.4144076\\
TUR & $\mathcal{M}_{\alpha^+}$ & 0.5613733 & 0.0036904 & 0.9221724 & 1006.5304764 & -455.9797216\\
TUR & $\mathcal{M}_{\alpha^-}$ & 0.5743223 & 0.0000163 & 0.9212163 & 1004.4008375 & -452.2555678\\
TUR & $\mathcal{M}_{\delta}$ & 0.5537710 & 0.0000113 & 0.9241992 & 1010.4413593 & -464.0276218\\
TUR & $\mathcal{M}_{\mu}$ & 0.5564595 & 0.0000000 & 0.9225319 & 1006.5641070 & -457.3917962\\
TUR & $\mathcal{M}_{null}$ & 0.5743230 & 0.0000163 & 0.9212166 & 1004.4010786 & -457.9771897\\
UKR & $\mathcal{M}_{\alpha^+}$ & 0.2824879 & 0.0000003 & 0.8475939 & 237.3301802 & -170.0780074\\
UKR & $\mathcal{M}_{\alpha^-}$ & 0.2448255 & 0.0000002 & 0.8612119 & 241.8217522 & -181.3100499\\
UKR & $\mathcal{M}_{\delta}$ & 0.1863009 & 0.0000001 & 0.8928971 & 257.3043951 & -212.4090202\\
UKR & $\mathcal{M}_{\mu}$ & 0.2496387 & 0.0000212 & 0.8602308 & 242.1071349 & -180.4647504\\
UKR & $\mathcal{M}_{null}$ & 0.2824878 & 0.0000003 & 0.8476001 & 237.3437052 & -174.8704237\\
URY & $\mathcal{M}_{\alpha^+}$ & 0.3707057 & 0.0004125 & 0.9539609 & 760.9348292 & -446.7383706\\
URY & $\mathcal{M}_{\alpha^-}$ & 0.3801091 & 0.0000909 & 0.9515936 & 754.4475193 & -433.2002436\\
URY & $\mathcal{M}_{\delta}$ & 0.3801091 & 0.0739791 & 0.9515934 & 754.4467727 & -433.1992421\\
URY & $\mathcal{M}_{\mu}$ & 0.4148870 & 0.0000000 & 0.9462801 & 739.8061158 & -405.0794658\\
URY & $\mathcal{M}_{null}$ & 0.3801091 & 0.0000489 & 0.9515934 & 754.4502665 & -438.7974684\\
USA & $\mathcal{M}_{\alpha^+}$ & 0.1668963 & 0.0055876 & 0.9697761 & 1854.4974683 & -1504.7078129\\
USA & $\mathcal{M}_{\alpha^-}$ & 0.1691152 & 0.0000027 & 0.9698283 & 1854.8253141 & -1505.6582104\\
USA & $\mathcal{M}_{\delta}$ & 0.1869925 & 0.0006049 & 0.9658160 & 1820.5736973 & -1437.1131786\\
USA & $\mathcal{M}_{\mu}$ & 0.1691152 & 0.0211774 & 0.9698284 & 1854.8269747 & -1505.6591624\\
USA & $\mathcal{M}_{null}$ & 0.1691152 & 0.0221088 & 0.9698284 & 1854.8267976 & -1511.9671769\\
VEN & $\mathcal{M}_{\alpha^+}$ & 0.7248894 & 0.0302883 & 0.9314021 & 368.7917156 & -158.4962726\\
VEN & $\mathcal{M}_{\alpha^-}$ & 0.6719343 & 0.0000000 & 0.9369305 & 373.8565502 & -168.7471573\\
VEN & $\mathcal{M}_{\delta}$ & 0.7248895 & 0.0000351 & 0.9314017 & 368.7906482 & -158.4955756\\
VEN & $\mathcal{M}_{\mu}$ & 0.7248898 & 0.0371293 & 0.9314029 & 368.7920561 & -158.4976122\\
VEN & $\mathcal{M}_{null}$ & 0.7248892 & 0.0000001 & 0.9314027 & 368.7897007 & -163.3013450\\
VNM & $\mathcal{M}_{\alpha^+}$ & 2.3967138 & 0.0012084 & 0.5717167 & 960.8749519 & -14.6593163\\
VNM & $\mathcal{M}_{\alpha^-}$ & 2.4133088 & 0.0000400 & 0.5678732 & 960.0251253 & -13.0243703\\
VNM & $\mathcal{M}_{\delta}$ & 2.4132716 & 0.0000647 & 0.5679233 & 959.9523137 & -13.0455685\\
VNM & $\mathcal{M}_{\mu}$ & 2.2242733 & 0.0000015 & 0.5854469 & 961.8716340 & -20.6221612\\
VNM & $\mathcal{M}_{null}$ & 2.4132930 & 0.0000029 & 0.5678849 & 959.9594100 & -18.2388068\\
ZAF & $\mathcal{M}_{\alpha^+}$ & 0.4738056 & 0.0014702 & 0.9471435 & 1054.7190010 & -621.0423136\\
ZAF & $\mathcal{M}_{\alpha^-}$ & 0.4936819 & 0.0000042 & 0.9466293 & 1051.1700514 & -617.4985641\\
ZAF & $\mathcal{M}_{\delta}$ & 0.4936820 & 0.0001680 & 0.9466294 & 1051.1688630 & -617.4996371\\
ZAF & $\mathcal{M}_{\mu}$ & 0.4936819 & 0.0000044 & 0.9466295 & 1051.1689693 & -617.4999436\\
ZAF & $\mathcal{M}_{null}$ & 0.4936819 & 0.0001973 & 0.9466293 & 1051.1698206 & -623.4012024
\end{longtable}

\newpage

\begin{table}[ht]
\centering
    \begin{tabular}{|l||r|r|}
\hline
                    & \multicolumn{1}{c|}{Log-likelihood} & \multicolumn{1}{c|}{Log-likelihood residuals} \\ 
\hline
\hline
(Intercept)         &                             0.000 &                                     1.495** \\ 
                    &                           (0.045) &                                     (0.474) \\
NT                 &                          0.928*** &                                             \\ 
                    &                           (0.046) &                                             \\
N                   &                                   &                                    -0.230** \\ 
                    &                                   &                                     (0.071) \\
log(GDP)            &                                   &                                    0.693*** \\ 
                    &                                   &                                     (0.187) \\
\hline
\hline
Number of countries &                                69 &                                          69 \\
\hline
$R^2$               &                             0.861 &                                       0.174 \\ 
\hline
\end{tabular}
\caption{\textbf{Linear regression model coefficients for the effect of the number of data points ($NT$) on the best model log-likelihood (middle column), and linear regression model coefficients for the effects of gross domestic product (GDP) and number of economic activities ($N$) on the residuals (right column)}. $^{**} p<0.01$, $^{***} p < 0.001$.}\label{table:gdp_loglikelihood}
\end{table}
\FloatBarrier
\end{appendices}

\end{document}